\documentclass[aps,preprint,nofootinbib,preprintnumbers,eqsecnum,superscriptaddress]{revtex4-1}

\usepackage{color}

\newcommand{\hl}[1]{\color{magenta}}

\usepackage[
	  pagebackref=false,
	  colorlinks=true,
      linkcolor=blue,
      urlcolor=blue,
      filecolor=black,
      citecolor=red,
      pdfstartview=FitV,
      pdftitle={},
        pdfauthor={},
        pdfsubject={},
        pdfkeywords={},
        pdfpagemode=None,
        bookmarksopen=true
      ]{hyperref}

\usepackage[normalem]{ulem}
\usepackage{amsmath,bm}
\usepackage[dvipsnames]{xcolor}
\usepackage{enumerate}
\usepackage{amsfonts}
\usepackage{epsfig}
\usepackage{mathbbol}
\usepackage{leftidx}
\usepackage[utf8]{inputenc}

\newcommand\half{{\ensuremath{\frac{1}{2}}}}

\newcommand{\be}{\begin{equation}}
\newcommand{\ee}{\end{equation}}
\newcommand{\bea}{\begin{eqnarray}}
\newcommand{\eea}{\end{eqnarray}}
\newcommand{\bega}{\begin{gather}}
\newcommand{\eega}{\end{gather}}

\newcommand{\bi}{\begin{itemize}}
\newcommand{\ei}{\end{itemize}}
\newcommand{\ben}{\begin{enumerate}}
\newcommand{\een}{\end{enumerate}}
\newcommand{\bca}{\begin{cases}}
\newcommand{\eca}{\end{cases}}
\newcommand{\bln}{\begin{align}}
\newcommand{\eln}{\end{align}}
\newcommand{\bst}{\begin{split}}
\newcommand{\est}{\end{split}}
\def\ie{\begin{equation}\begin{aligned}}
\def\fe{\end{aligned}\end{equation}}
\newcommand{\bma}{\le(\begin{matrix}}
\newcommand{\ema}{\end{matrix}\ri)}

\newcommand\al{{\alpha}}
\newcommand\ep{\epsilon}
\newcommand\sig{\sigma}
\newcommand\Sig{\Sigma}
\newcommand\lam{\lambda}
\newcommand\Lam{\Lambda}

\newcommand\Om{\Omega}

\newcommand\De{{\ensuremath{{\Delta}}}}

\newcommand\ov{\over}
\newcommand\ha{{\half}}

\def\le{\left}
\def\ri{\right}

\newcommand{\bde}{\bm{\delta}}

\begin{document}

\title{Perturbation theory for the logarithm of a positive operator}

\preprint{MIT-CTP/5081}

\author{Nima Lashkari}
\affiliation{School of Natural Sciences, Institute for Advanced Study, Einstein Drive, Princeton, NJ, 08540}
\vspace*{1cm}
{\let\thefootnote\relax\footnote{$\mathrm{lashkari@ias.edu}$, $\mathrm{hong\_liu@mit.edu}$, $\mathrm{srivat91@mit.edu}$.}}

\author{Hong Liu}
\affiliation{Center for Theoretical Physics, \\
Massachusetts Institute of Technology, Cambridge, MA 02139 }

\author{Srivatsan Rajagopal}
\affiliation{Center for Theoretical Physics, \\
Massachusetts
Institute of Technology,
Cambridge, MA 02139 }

\begin{abstract}
\noindent 

In various contexts in mathematical physics one needs to compute the logarithm of a positive unbounded operator. 
Examples include the von Neumann entropy of a density matrix and the flow of operators with the modular Hamiltonian in the Tomita-Takesaki theory. 
Often, one encounters the situation where the operator under consideration, that we denote by $\Delta$, can be related by a perturbative series to another operator $\Delta_0$, whose logarithm is known. 
We set up a perturbation theory for the logarithm $\log \De$.  It turns out that the terms in the series possess
remarkable algebraic structure, which enable us to write them in the form of nested commutators plus some 
``contact terms."

\end{abstract}

\today

\maketitle

\tableofcontents

\section{Introduction}

In many different problems in mathematical physics one needs to compute the logarithm of a positive operator. 
This is commonplace in asymptotic quantum information theory when one is interested in various quantities constructed from the logarithm of a density matrix. For instance, given a reduced density matrix $\rho$, one needs to compute $\log \rho$ to find the von Neumann entropy and relative entropies. 
In a general quantum system, in the Tomita-Takesaki theory, given the modular operator $\Delta_\Om$ of a state $|\Om\rangle$ or the relative modular operator of two states $\Delta_{\Psi\Om}$ one needs to compute their logarithms to obtain the modular flow operator $\De_\Om^{it}$, or to calculate relative entropies. In this case, the positive operator in question, $\Delta_\Om$, is unbounded. 

Consider an unbounded positive operator $\De$. In general, obtaining $\log \De$ directly is difficult since it has a simple form only in the spectral decomposition of the operator $\Delta$. 
We consider the following situation: (i) $\De$ is related to some other positive operator $\De_0$ by a smooth deformation, i.e. there exists a continuous parameter $\lam$ and a family of operators $\De (\lam)$ that interpolate between $\De (0) = \De_0$ and $\De(1) = \De$; (ii) 
the logarithm $\log \De_0$ is known explicitly. 
Imagine setting up a perturbative series for $\log \De (\lam)$ in terms of $\log \De_0$ for $\lam$ small. If the perturbation series converges for $\lam\leq 1$ one can extend the series to $\lam =1$. 
It is the goal of this paper to set up such a perturbation theory. For a discussion of the fractional powers and the logarithm of bounded operators in the Hilbert space see ~\cite{Komatsu:1966,logarithmop:1987}. For bounded operators that belong to the Lie algebra of a Lie group one often uses the Baker-Campbell-Hausdorff (BCH) expansion to compute the logarithm; see~\cite{Liealgebrabook:2013}. See also recent discussions~\cite{Faulkner:1,Faulkner:2,Faulkner:3,Faulkner:4, Faulkner:5, Sarosi:2017} in the context of quantum field theory. 

Our main result is the following series expansion
\be \label{nju}
\log \De_0 - \log \De = \sum_{m=1}^\infty Q_m
\ee
where $Q_m$ can be written in the form of nested commutators plus ``contact'' contributions
\begin{align}
Q_m & = {2 \pi \ov m} \text{lim}_{\epsilon_i\rightarrow 0}\int dt_1\int dt_2...\int dt_m \, F_{\epsilon_i}(t_1,t_2,...,t_m)[\cdots [{\bm{\delta}}(t_1),{\bm{\delta}(t_2)}], \cdots], {\bm{\delta}}(t_m)]+P_m
\label{intrep0}
\end{align}
and 
\bega
\bm{{\delta}}(t) =\Delta^{-it}_0 \bm{{\delta}}\Delta^{it}_0 , \qquad  {\bm{\delta}}=\frac{\bm{\al}}{1-\bm{\al}/2} , \qquad  \bm{\al}= 1-\Delta^{-\ha}_0\Delta \Delta^{-\ha}_0,  \\
\label{defF}
F_{\epsilon_i}(t_1,t_2,..,t_m)= f(t_1)g_{\epsilon_1}(t_2-t_1)g_{\epsilon_2}(t_3-t_2)...g_{\epsilon_{m-1}}(t_{m}-t_{m-1})f(t_m)  , \\
f(t)=\frac{1}{2\text{cosh}(\pi t)}, \qquad
g_{\epsilon}(t)=\frac{i}{4}\biggl[\frac{1}{\text{sinh}(\pi(t-i\epsilon))}+\frac{1}{\text{sinh}(\pi(t+i\epsilon))}\biggr] \ .
\end{gather}
In~\eqref{intrep0} $P_m$ are given by terms with two fewer integrals (``contact terms'') whose structure are a bit complicated and 
will be given later. 
 The first few terms of this series are given by 
\bea
&&Q_1 = {\pi \ov 2} \int_{-\infty}^\infty  \frac{dt}{\cosh^2(\pi t)}   \bde (t) , \label{firsteq}\\
&&Q_2 =  {\pi \ov 4} \lim_{\ep\to 0}  \int_{-\infty}^\infty \frac{dt_1 dt_2}{\cosh(\pi t_1)\cosh(\pi t_2)}g_\ep (t_2-t_1)\left[ \bde (t_1),   \bde (t_2) \right]  , \label{secondeq}\\
&&Q_3 =  {\pi \ov 6} \lim_{\ep_1, \ep_2 \to 0} \int \frac{dt_1 dt_2dt_3}{\cosh(\pi t_1)\cosh(\pi t_3)}g_{\ep_1} (t_2-t_1)g_{\ep_2} (t_3-t_2) \,  [[\bde (t_1),\bde(t_2)],  \bde (t_3)] \nonumber\\
&&+\frac{\pi}{24}\int \frac{\bde(t)^3}{\text{cosh}^2(\pi t)} \label{thirdeq}\\
&&Q_4 = \lim_{\ep_1,\ep_2,\ep_3\to 0} \frac{\pi}{8} \int 
\frac{dt_1 dt_2dt_3 dt_4 \, g_{\ep_1} (t_2-t_1)g_{\ep_2} (t_3-t_2) g_{\ep_3} (t_4-t_3)}{\cosh(\pi t_1)\cosh(\pi t_4)}
[[[\bde(t_1),\bde(t_2)],\bde(t_3)],\bde(t_4)] \nonumber\\
&&+ \frac{\pi}{32}\lim_{\ep\to 0} \int \frac{dt_1 dt_2 \, g_\ep (t_2-t_1) }{\cosh(\pi t_1)\cosh(\pi t_2)} \left\lbrace\bde(t_2),[\bde(t_1),\bde^2 (t_2)]\right\rbrace\label{fourtheq}
\eea
It is important in~\eqref{intrep0} that performs the integrals keeping $\ep$'s nonzero and then take the $\ep_i \to 0$ limit.  We have included quintic contact term $P_5$ in Appendix~\ref{sec:quintic}.

In the special case the operators $\Delta$ and $\Delta_0$ are both bounded one can use the spectral representation of these operators to match our expansion and the BCH expansion order by order in $\lambda$.

The plan of the paper is as follows. In Sec.~\ref{sec:out}, we outline the main steps leading to the proof of~\eqref{intrep0} and give explicit expressions for $F_m$.  In Sec.~\ref{sec:Pf1} and Sec.~\ref{sec:Pf3} we fill in the details of the proof. In Appendix~\ref{app:b} we give a simple example of harmonic oscillator to illustrate the use of~\eqref{firsteq}--\eqref{fourtheq}. In Appendix~\ref{sec:quintic} we present the explicit expression for $P_{m=5}$. Appendices~\ref{app:a}--\ref{sec:Pf5} include various fine details for the proof. 

\section{Outline of the proof} \label{sec:out}

%In this section we outline the main ingredients for the proof of~\eqref{intrep0}.  

\subsection{Setup} 

We start with the integral representation of the logarithm of an (positive invertible) operator $\Delta$:
\begin{align}
\log \Delta &=\int_0^\infty d\beta\biggl(\frac{1}{1+\beta}-\frac{1}{\Delta+\beta}\biggr)\ .
\end{align}
For unbounded operators $\Delta$, the integral on the right-hand-side should be thought of as a limit of Riemann sums in the strong operator topology induced by the domain of the logarithm of $\Delta$. Thus, we have the operator equality
\begin{align}
\text{log}\Delta_0 -\text{log}{\Delta}&=\int_0^\infty d\beta\biggl(\frac{1}{{\Delta}+\beta}-\frac{1}{\Delta_0+\beta}\biggr) =\int_0^\infty d\beta\frac{1}{\Delta_0+\beta}(\Delta_0-{\Delta})\frac{1}{{\Delta}+\beta} \ 
\label{intrep}
\end{align}
on the common domain of $\text{log}\Delta_0$ and $\text{log}{\Delta}$.

Introduce the operator
\begin{align}
\bm{\al}&= 1-\Delta^{-\ha}_0 {\Delta}\Delta_0^{-\ha}
\end{align}
which we take to depend on a continuous  parameter $\lam$ and vanish as $\lam \to 0$. To lighten the notation, we keep the $\lam$ dependence implicit. We stress that $\bm{\al}$ is an unbounded operator despite the fact that it is proportional to a small parameter $\lam$. 
%\NL{Comment on operator distance for unbounded operators?}

Since the function $f(x)=\sqrt{x}(x+\beta)^{-1}$ is bounded for positive $\beta$ and $x$ the operator $\Delta^{1/2}(\Delta+\beta)^{-1}$ is a bounded operator in the Hilbert space. To make sense of the perturbation theory we assume that there exists a constant $c$ such that $\Delta(\lam)<c\Delta_0$. As a result, the operator $\Delta_0^{1/2}(\Delta+\beta)^{-1}$ is also bounded. This is part of what we mean by $\lam$ being a small perturbation.

For $\beta>0$, we define the bounded operator
\bea
A=\frac{\Delta_0^{\ha}}{\Delta_0+\beta}
\eea
to rewrite the integrand of Eq.(\ref{intrep}) as
%Through a few lines of algebraic manipulations (each of which is justified since for $\beta>0$, all involved operators are bounded), we find that the integrand of Eq.(\ref{intrep}) can be written as 
\bea\label{jui}
&&\frac{1}{\Delta+\beta}-\frac{1}{\Delta_0+\beta}=\frac{1}{\Delta_0+\beta}(\Delta_0-{\Delta})\frac{1}{{\Delta}+\beta}\\
&&=A\bm{\al} \Delta_0^{1/2}(\Delta+\beta)^{-1}= A\bm{\alpha}(1-\Delta_0^{1/2}A\bm{\alpha})^{-1}A\\
&&=A{\bm{\delta}}(1-B{\bm{\delta}})^{-1}A,
\eea
where we have introduced 
\be
{\bm{\delta}}=\frac{\bm{\al}}{1-\bm{\al}/2}, \qquad B =\frac{\Delta_0-\beta}{2(\Delta_0+\beta)} \ .
\ee

One might want to naively expand $(1-\Delta_0^{\ha}A\bm{\al})^{-1}$ in the second line of (\ref{jui}) in a power series of $\Delta_0^{\ha}A\bm{\al}$. But this is not a good expansion as  $\Delta_0^{\ha}A\bm{\al}$ is an unbounded operator. 
This is similar the approach taken by \cite{Sarosi:2017}. It leads to singular integrals which can be sensible only if one provides a prescription to deform the integration Contour.
To  circumvent this problem, in the third line of~\eqref{jui} we introduced the operator ${\bm{\delta}}$, which is bounded  with a norm $\| \bde \|\leq 2$. To see this note that the spectrum of the closure of $\bm{\al}$ is contained in $(-\infty,1)$ and $\frac{x}{1-x/2}$ is a bounded function in the range $(-2,2)$. On a dense domain, $\bm{\delta}$ and its closure agree. Finally, expanding $B$ in the spectral decomposition of $\Delta_0$ we find that the spectrum of $B$ is contained in $\left( 0,\frac{1}{2}\right)$. Therefore, $||B||=1/2$ and by the Cauchy-Schwarz inequality $||B\bm{\delta}||\leq 1$.

 For $||B\bm{\delta}|| < 1$ expanding the third line of ~\eqref{jui} in terms of $B \bde$ gives a convergent series. In general, it is not possible to exclude the $||B\bm{\delta}||=1$ case.
To justify our expansion, we will restrict to those vectors $|x\rangle$ in Hilbert space which satisfy 
\begin{align}
||B\bm{\delta} A|x\rangle|| < ||A|x\rangle||\ .
\end{align}
On this set, we have the operator equality
\begin{align}
A\bm{\delta}(1-B {\bm{\delta}})^{-1}A |x\rangle=A\bm{\delta}\sum_{n=0}^\infty (B {\bm{\delta}})^nA |x\rangle  \ .
\label{geometricsum}
\end{align}
and the sum on the right-hand-side is pointwise convergent. We will not specify $|x\rangle$ below but its presence should always be kept in mind.

Using~\eqref{jui}--\eqref{geometricsum} in~\eqref{intrep} we then find that 
\begin{align}
\text{log}\Delta_0-\text{log} {\Delta}&=\int_0^\infty d \beta A {\bm{\delta}}\sum_{n=0}^\infty (B {\bm{\delta}})^n A \ .\label{geometricsum2}
\end{align}
We now further rewrite the above expression using the one-parameter unitary group $\De_0^{it}$ generated by $\log \Delta_0$. In particular, we use the following integral expressions for $A$ and $B$ 
\begin{align} \label{intrep3}
A %=\frac{\Delta^{1/2}}{\Delta+\beta}
=\frac{1}{\sqrt{\beta}}\int_{-\infty}^\infty dt \, f(t)\beta^{it}\Delta^{-it}_0, \qquad 
B =\text{lim}_{\epsilon\rightarrow 0}\int_{-\infty}^\infty dt \, g_{\epsilon}(t)\beta^{it}\Delta^{-it}_0 
\end{align} 
with 
\be
f(t)=\frac{1}{2\text{cosh}(\pi t)}, \qquad 
g_{\epsilon}(t) =\frac{i}{4}\biggl[\frac{1}{\text{sinh}(\pi(t-i\epsilon))}+\frac{1}{\text{sinh}(\pi(t+i\epsilon))}\biggr] \ .
\ee
In Appendix~\ref{sec:Pf4}, we show that the above $\ep\to 0$ limit exists. 

Plugging~\eqref{intrep3} into~\eqref{geometricsum2} we find that
\bea \label{ejn}
\log  \De_0 - \log  \De & =&  \sum_{m=0}^\infty\int_0^\infty {d\beta \ov \beta} \int_{-\infty}^\infty dt_0 \cdots dt_{m+1}\,
\beta^{i(t_0+\cdots +t_{m+1})}  \cr
&& \times \; f(t_0)  \Delta^{-i t_0}_0 \bde  \le(\prod_{i=1}^m g_{\ep_i}  (t_i)  \Delta_0^{-i t_i} \bde \ri)  \Delta_0^{-i t_{m+1}} f(t_{m+1}) \ .
% \Delta^{i t_1}\cdots \Delta^{i t_{m+1}}
\eea
%The reason we chose to use $\tilde \de$ and $B$ through the second equality in~\eqref{euq}, instead of the expression 
%in terms of $\de$ and $A$, is that with more than two factors of $A$, the $\beta$ integral in~\eqref{ejn} would not be well defined due to too many factors of $\beta^{-\ha}$. 
Notice that if we exchange the orders of $\beta$ and $t$-integrals (with associated $\ep_i \to 0$ limit) the $\beta$-integral can be performed explicitly 
\be
\int_0^\infty \beta^{-1}d\beta\: \beta^{ i(t_0+\cdots+t_{m+1})}=2\pi\delta(t_0+\cdots +t_{m+1})\ .
\ee
Equation~\eqref{ejn} can then be further written as (shifting the sum of $m$ to start from $1$)
\be \label{poi}
\log \De_0 - \log \De = \sum_{m=1}^\infty Q_m
\ee
with 
\bega \label{poi1}
Q_m = 2 \pi \, \text{lim}_{\epsilon_i\rightarrow 0}\int dt_1\int dt_2...\int dt_m \, F_{\epsilon_i}(t_1,t_2,\cdots,t_m) \, {\bm{\delta}}(t_1) \cdots {\bm{\delta}}(t_m) 
\end{gather}
where the kernel $F$ is defined by
\begin{align}\label{poi2}
F_{\epsilon_i}(t_1,t_2,..,t_m)&=f(t_1)g_{\epsilon_1}(t_2-t_1)g_{\epsilon_2}(t_3-t_2)...g_{\epsilon_{m-1}}(t_{m}-t_{m-1})f(t_m) 
\end{align}
and 
\be
\bm{{\delta}}(t) =\Delta^{-it}_0 \bm{{\delta}}\Delta^{it}_0  \ .
\ee
A variant of~\eqref{poi}--\eqref{poi2} has appeared previously in~\cite{Sarosi:2017}.\footnote{Note that the expansion in~\cite{Sarosi:2017} is similar to expanding (\ref{jui}) in $\alpha$ which is an unbounded operator in our case. To have a convergent series we use the bounded operator $\bm{\delta}$ as the expansion parameter. Our expressions~\eqref{poi}-\eqref{poi2} 
appear to differ from that in~\cite{Sarosi:2017} in the $i \ep$ prescription, the integration contours, and operator orderings.} The main goal of the paper is to show that the kernel~\eqref{poi2} has remarkable symmetric properties 
which enable one to write $Q_m$ in terms of nested commutators of $\bde$'s 
\begin{align} \label{iu}
Q_m %&=\text{lim}_{\epsilon_i\rightarrow 0}\int dt_1\int dt_2...\int dt_m F_{\epsilon_i}(t_1,t_2,...,t_m){\bm{\delta}}(t_1)...{\bm{\delta}}(t_m) \\
& = {2 \pi \ov m} \text{lim}_{\epsilon_i\rightarrow 0}\int dt_1\int dt_2...\int dt_m \, F_{\epsilon_i}(t_1,t_2,...,t_m)[\cdots [{\bm{\delta}}(t_1),{\bm{\delta}(t_2)}], \cdots], {\bm{\delta}}(t_m)]+P_m \ .
\end{align}
 The $P_m$ contact term involves terms which contain two fewer integrals. It has the following structure:
 \be 
 P_m = \sum_{s} \int dt_1 \cdots dt_{m-3} dt \, \, J_s (t_1, \cdots t_{m-3}; t) M_s (t_1, \cdots t_{m-3}; t)\label{iu2},
 \ee
where $s$ sums over all possible ways in which three $t_i$'s are selected from the set $\{t_1, \cdots t_m\}$ such that at least two of the indices on the chosen $t_i$ are adjacent. The three chosen $t_i$'s are set to be equal to $t$, with the rest relabeled as $t_1 , \cdots t_{m-3}$. $J_s$ is a kernel which 
can be obtained from $F_{\ep_i}$ after applying a number of operations which are described in the next subsection. For now, we give some simple examples.  Suppose $t_1 = t_2 =t_3 = t$, then
\begin{align}
M=\delta(t)^3\delta(t_1)...\delta(t_{m-3}), \qquad
J=\frac{\pi}{2m}F_{\epsilon}(t,t_1,...,t_{m-3})\ .
\end{align}
For $t_1=t_2=t_5=t$, one has  $M= \delta(t)^2\delta(t_1)\delta(t_2)\delta(t)\delta(t_3)...\delta(t_{m-3})$ and
\begin{align}
J=\frac{\pi}{2m}(F_{\epsilon}(t_2,t_1,t,t_3,t_4...,t_{m-3})-F_{\epsilon}(t_1,t_2,t,t_3,t_4,..,t_{m-3})) \ .
\end{align}
As the selected indices become larger, the number of terms in $J$ increases and the terms also become more complicated. For $t_6=t_2=t_3 =t$ one has $M=\delta(t_1)\delta(t)^2\delta(t_2)\delta(t_3)\delta(t)\delta(t_4)..\delta(t_{m-3})$, and $J$ has a term of the form $\frac{1}{f(t)^2}F(t_3,t_2,t,t_1)F(t,t_4,t_5,..,t_{m-3})$.

\subsection{Basic ideas for the proof}

%\section{Strategy for the Proof of the Nested commutator relation, Eq.(\ref{NestedComm})}
%\subsection{Basic Idea}

The kernel~\eqref{poi2} looks complicated, but it satisfies a number of amazing identities under the permutation of its arguments. 
To explain the basic idea leading to the proof of~\eqref{iu}, 
%is to show that the kernel~\eqref{poi2} acts as a projector onto the subspace of nested commutators of the $\bde(t)$.
we need to first establish some notation. 

Let $S_m$ be the symmetric group of permutations of $m$-distinct objects. We use the \textit{cycle} notation $(12 \cdots m)$ to represent the permutation that sends $1\rightarrow 2\rightarrow 3 \cdots \to (m-1)\rightarrow  m\rightarrow 1$. Any index not listed in the cycle is left untouched. 
We define the action of an element $\sig  \in S_m$ on a product of operators $\bde(t_{i_1})\bde(t_{i_2})..\bde(t_{i_m})$ and a general function $H(t_{i_1},t_{i_2},..,t_{i_m})$ in the following manner:
\begin{align}\label{i1}
\sigma(\bde(t_{i_1})\bde(t_{i_2})..\bde(t_{i_m}))&:= \bde(t_{i_{\sigma (1)}})\bde(t_{i_{\sig(2)}})..\bde(t_{i_{\sig(m)}})\\
\sigma(H(t_{i_1},t_{i_2},...,t_{i_m}))&:= H(t_{\sigma{(i_1)}},t_{\sigma(i_2)},..,t_{\sigma(i_m)})
\label{i2}
\end{align}
%where in~\eqref{i1}, $1\sigma$ means the action of $\sigma$ on $1$ and so on. 
Note that in~\eqref{i1} $\sig$ acts on the left while in~\eqref{i2} it acts on the right. 
See Appendix~\ref{app:a} for further explanations and examples.  Let us also introduce the special permutations
\begin{align}
\mu_j  = (j(j-1)(j-2)....4321), \quad 
\Lambda_j = (1234...(j-1)j), \quad \mu_j = \Lam_j^{-1} \ .
\end{align}

One can show that the following statements are true: 

\ben 

\item Introduce the operator
\begin{align}
T_m &= (\text{id}-\mu_m)(\text{id}-\mu_{m-1})...(\text{id}-\mu_2)
\end{align} 
where ${\rm id}$ denotes the identity operation. Then, 
\begin{align}
T_m \bigl(\bde(t_1)...\bde(t_m)\bigr) = [[\cdots [{\bm{\delta}}(t_1),{\bm{\delta}(t_2)}], \cdots], \bde (t_{m-1})], {\bm{\delta}}(t_m)]  \ .
%\text{NestedCommutator}[\bde(t_1),\bde(t_2),...,\bde(t_m)]
\label{Nestcomm 1}
\end{align}
See~\cite{Blessenohl:1988} for a proof. For completeness, we have included a proof in Appendix~\ref{app:a}.

\item For any function $H(t_1,t_2,..,t_m)$ we have
 \begin{align} \label{Nestcomm 2}
\int H(t_1,t_2,..,t_m) \bde(t_1)\bde(t_2)..\bde(t_m)&=\int H(t_1,t_2,..,t_m) \, T_m\bigl(\bde(t_1)\bde(t_2)..\bde(t_m)\bigr)
\cr
&+\int \bde(t_1)\bde(t_2)..\bde(t_m)\, \Sigma_m H(t_1,t_2,..,t_m)
\end{align}
where the operation $\Sig_m$ is defined as 
\be
\Sigma_m  H(t_1,t_2,..,t_m)  =\sum_{k=1}^{m-1}(-1)^{k-1} \, \Xi_k^m H(t_1,t_2,..,t_m)
\label{Nestcomm 4}
\ee
with
\be
\Xi_k^m H(t_1,t_2,..,t_m) \equiv \sum_{i_1=k+1}^m\sum_{i_2=k}^{i_1-1}\sum_{i_3=k-1}^{i_2-1}...\sum_{i_k=2}^{i_{k-1}-1}\Lambda_{i_1}\Lambda_{i_2}...\Lambda_{i_k}H(t_1,t_2,..,t_m) \label{bbu}\ .
\ee

\item For a general operator $O(t_1,..,t_m)$ and function $F_\ep$ of~\eqref{poi2} we have
\be\label{pou}
\lim_{\ep_i \to 0} \int O(t_1,..,t_m) \, \Xi_k^m F_\epsilon(t_1,..,t_m)
= (-1)^k \lim_{\ep_i \to 0} \int O(t_1,..,t_m) \,  F_\epsilon(t_1,..,t_m) +\sum_{j=k+2}^m N_j[O] 
\ee
where $N_j[O] $ are ``contact terms'' containing only $m-2$ integrals. They will be given explicitly at the end. 

\item If we set $H = F_\ep$ in~\eqref{Nestcomm 2}  and use ~\eqref{pou} in~\eqref{Nestcomm 4} we find that 
\bega
Q_m = {2 \pi \ov m} 
\lim_{\ep_i \to 0}  \int F_\epsilon (t_1,t_2,..,t_m) \: T_m\biggl(\bde(t_1)\bde(t_2)..\bde(t_m)\biggr) 
+\frac{2 \pi}{m}\sum_{k=1}^{m-1}(-1)^{k-1}\sum_{j=k+2}^m N_j
\end{gather}
which leads to~\eqref{iu} upon using~\eqref{Nestcomm 1} together with the identification 
\begin{align}
P_m = \frac{2\pi}{m}\sum_{k=1}^{m-1}(-1)^{k-1}\sum_{j=k+2}^m N_j  \ .
\end{align}
\een

We now give the explicit expressions for $N_j$. First, let us introduce some definitions. For integers $q_2<q_1$
we define
%need to define the objects $R^{(q_1)}_{q_2}[M]$ for some integer $q_2<q_1$. It is given by  
%The distribution $R_{j}^{(q)}[M]$ is given by
\begin{align}
L^\epsilon_{q_1,q_2}(t)& \equiv \frac{\Lambda_{q_2} F_{\epsilon}(t_1,...,t_{q_1})}{g_{\epsilon}(t_1-t_{q_2})g_{\epsilon}(t_{q_2+1}-t_1)}\biggr\vert_{t_1=t_{q_2}=t_{q_2+1}=t}, \label{contact1}\\
R_{q_2}^{(q_1)}[M]& \equiv \text{lim}_{\epsilon_i\rightarrow 0}\int L^\epsilon_{q_1,q_2}(t)M(t_1,t_2,..,t_{q_1})\vert_{t_1=t_{q_2}=t_{q_2+1}=t}  \label{contact2}
\end{align}
where $M$ is some operator and on the right hand side of~\eqref{contact2} the integrations are over all distinct $t_i$'s (i.e. $q_1-2$ integrations). %Moreover, the index $q_2$ in $R^{(q_1)}_{q_2}$ instructs us as to which variables are set equal to one another. 
Given an operator $O(t_1,t_2,..,t_m)$, for each $p\leq k$ and $k+3\leq j\leq m$ where $k$ is the lower index in the object $\Xi^m_k$, we define the operator $W$ as
\begin{align}
&W(t_1,t_2,...,t_p,t_{j-1},t_{j}) =\frac{1}{f(t_j)f(t_{j-1})}\sum_{i_{p+1}=k-p+1}^{j-p-2}\sum_{i_{p+2}=k-p}^{i_{p+1}-1}...\sum_{i_k=2}^{i_{k-1}-1}\label{EQ1}\cr &  %\text{lim}_{\epsilon\rightarrow 0}
\int \biggl( \Pi_{n=p+1}^{j-2}dt_n \Pi_{l=j+1}^m dt_l\biggr) \, \chi_{pj}  O(t_1,...t_m)g_{j,j+1,\epsilon_j}g_{j+1,j+2,\epsilon_{j+1}}..g_{m-1,m,\epsilon_{m-1}}f_m  
\end{align}
where $\chi_{pj} $ is given by 
\begin{align}
\chi_{pj}(t_{p+1},t_{p+2},..,t_{j-2},t_{j-1})&= \Pi_{n=j-1}^{j-p}\Lambda_n \Pi_{l=p+1}^k\Lambda_{i_{l}} \biggl(f_1g_{1,2,\epsilon_1}g_{2,3,\epsilon_2}...g_{j-p-2,j-p-1,\epsilon_{j-p-2}}\biggr),
\label{EQ2}
\end{align}
and we have used the following shorthand notations
\begin{align}\label{shorthand}
f_m \equiv f(t_m), \quad g_{m-1,m,\epsilon_{m-1}} \equiv g_{\ep_{m-1}} (t_m - t_{m-1})\ .
\end{align}

We also introduce 
\be
\tilde{O}_j(y_1,y_2,...,y_{p+2})=W(y_{p+2},y_{p+1},...,y_3,y_2,y_1)\label{EQ3}
\ee
which amounts to the relabelling
\begin{align}
&y_1=t_j, \quad y_2=t_{j-1},  \quad 
y_{3+q}=t_{p-q},\hspace{20pt}0\leq q\leq p-1\label{Rel2}
\end{align}
Finally, the expression for $N_j[O]$ is given by
\begin{align}
N_j[O] = \bca  %\sum_{p=1}^k\sum_{l=1}^{p+2}\Lambda_l[\tilde{O}_j]=
\frac{1}{4}\sum_{p=1}^k\sum_{l=2}^{p+1}R^{(p+2)}_l[\tilde{O}_j]&j\geq k+3\label{Ans1}\cr
 %\sum_{l=1}^{k+2}\Lambda_l[\tilde{O}_j]=
 \frac{1}{4}\sum_{l=2}^{k+1}R^{(k+2)}_l[\tilde{O}_{k+2}]&j=k+2
 \eca
\end{align}
for $m \geq 2$ and $1 \leq k \leq m-1$.

In summary, to obtain the kernel $J$ of (\ref{iu2}) corresponding to $t_j, t_{r+1},t_{r}$ for some $1\leq r<j-1$, we use the following algorithm:  
\begin{enumerate}

\item Choose an integer $k$ such that $1\leq k\leq j-2$. If $k=j-2$ then set $p=k$, otherwise choose $p$ in the range $r+1\leq p\leq k$.

\item For each such $k$ and $p$, apply the string of permutations on the right-hand-side of Eq.(\ref{EQ2}) to the first $j-p-2$ arguments of $F_{\epsilon}$ to obtain $\chi_{pj}$. Sum over these permutations.
\item Keep all arguments above $j$ unpermuted, and delete the arguments between $j-p-2$ and $j$. Divide the result by $\frac{1}{f(t_j)f(t_{j-1})}$. 

\item Define the variables $y_i$ as in (\ref{Rel2}). Apply the permutation $\Lambda_{p+2-r}$ to $F(y_1,..,y_{p+2})$ and divide by $4g_{\epsilon}(y_1-y_{p+2-r})g_{\epsilon}(y_{p+3-r}-y_1)$. Now set $t_j=t_{r+1}=t_r$

\item Multiply this by the result obtained in Step 3 after reexpressing $y_i$ in terms of $t_i$. Perform the sum over $p$. Sum over $k$ after multipying by $(-1)^{k-1}$. Multiply the whole result by $\frac{2\pi}{m}$.
\end{enumerate}

To obtain $J$ corresponding to the choice $t_j=t_{j-1}=t_r$ when $1<r<j-2$, the previous steps are followed with the choice $p=r$ and no sum over $p$ in the last step. To obtain $J$ corresponding to the choice $t_j=t_{j-1}=t_1$ follow the previous steps after setting $p=1$ and to the result, add the term $\frac{\pi\Lambda_{j-1}F(t_1,..,t_m)}{2m(g_{\epsilon}(t_1-t_{j-1})g_{\epsilon}(t_j-t_1))}$. This exhausts all the cases.

The rest of the paper is devoted to establishing~\eqref{Nestcomm 1}--\eqref{pou} and justifying the existence of $\ep_i \to 0$ limit. 
In Sec.~\ref{sec:Pf1} we prove~\eqref{Nestcomm 1} and~\eqref{Nestcomm 2}. 
In Sec.~\ref{sec:Pf3} we prove~\eqref{pou}. Appendix~\ref{sec:Pf4} discusses in detail the $\ep_i \to 0$ limit. In Appendix~\ref{sec:Pf5} we examine more carefully the 
interchange of $\beta$-integral and $\ep_i \to 0$ limit used in~\eqref{ejn}. 

\section{Permutation identities (I)} \label{sec:Pf1}

In this section, we present a proof of Eq.~(\ref{Nestcomm 2}). Consider the integral
\begin{align}
I_m&=\int H(t_1,t_2,...,t_m)\bde(t_1)\bde(t_2)..\bde(t_m)
\end{align}
for some function $H(t_1,t_2,..,t_m)$. Then, we have
\begin{align}
I_m&= \int H(t_1,t_2,..,t_m)(1-(12)+(12))\bde(t_1)\bde(t_2)..\bde(t_m)\cr
&=\int H(t_1,t_2,..,t_m)\biggl[(1-(321))(1-(12))+(321)(1-(12))+(12)\biggr]\bde(t_1)\bde(t_2)..\bde(t_m) \cr
&=\int H(t_1,t_2,..,t_m)\biggl[(1-(4321))(1-(321))(1-(12))\cr
&+(4321)(1-(321))(1-(12))+(321)(1-(12))+(12)\biggr]\bde(t_1)\bde(t_2)..\bde(t_m) \ .
\label{pio}
\end{align}
Note that the third line  is simply $T_4\bigl[\bde(t_1)\bde(t_2) \bde (t_3) \bde(t_4) \big] \bde(t_5) \cdots \bde(t_m)$, while 
the fourth line when expanded gives
\begin{align}
\sum_{k=1}^{3}(-1)^{k-1}\sum_{i_1=k+1}^4\sum_{i_2=k}^{i_1-1}\sum_{i_3=k-1}^{i_2-1}...\sum_{i_k=2}^{i_{k-1}-1}\mu_{i_1}\mu_{i_2}...\mu_{i_k}\biggl(\bde(t_1)\bde(t_2)\bde(t_3)\bde(t_4)\biggr)  \bde(t_5) \cdots \bde(t_m) \ .
\end{align}
Continuing this process repeatedly, we get
\begin{align}
I_m&=\int H(t_1,t_2,..,t_m) \, T_m\biggl[\bde(t_1)\bde(t_2)..\bde(t_m)\biggr]
+\int H(t_1,t_2,..,t_m) \cr
&\times \sum_{k=1}^{m-1}(-1)^{k-1}\sum_{i_1=k+1}^m\sum_{i_2=k}^{i_1-1}\sum_{i_3=k-1}^{i_2-1}...\sum_{i_k=2}^{i_{k-1}-1}\mu_{i_1}\mu_{i_2}...\mu_{i_k}\biggl(\bde(t_1)\bde(t_2)..\bde(t_m)\biggr) \ .
\label{Lastbut2nd}
\end{align}
In Eq.~(\ref{Lastbut2nd}), the permutations act on the operators. But for subsequent applications, we need the permutations to act on functions. This is achieved by the obervations
\begin{align}
&\int H(t_1,t_2,...,t_m)\mu_{i_1}...\mu_{i_k}\biggl(\bde(t_1)..\bde(t_m)\biggr)\\&=\int H(t_1,t_2,...,t_m)\bde(t_{\tau(1)})\bde(t_{\tau (2)})...\bde(t_{\tau (m)})\hspace{20pt} \tau = \mu_{i_1}\star\mu_{i_2}\star...\star\mu_{i_k}\label{Product1}\\&=\int H(t_{\sigma(1)},t_{\sigma (2)}..t_{\sigma (m)})\bde(t_{\tau\sigma(1)})\bde(t_{\tau\sigma(2)})...\bde(t_{\tau\sigma(m)})\label{Perminvar}\\&=\int \tau^{-1}H(t_1,t_2,..,t_m)\bde(t_1)..\bde(t_m)\label{Conseq1}\\&=\int(\mu_{i_k}^{-1}\star\mu_{i_{k-1}}^{-1}...\mu_{i_2}^{-1}\star\mu_{i_1}^{-1})H(t_1,t_2,..,t_m)\bde(t_1)..\bde(t_m)\label{Conseq2}\\&=\int\biggl[\Lambda_{i_1}\Lambda_{i_2}..\Lambda_{i_k}H(t_1,t_2,..,t_m)\biggr]\bde(t_1)..\bde(t_m)\label{Conseq3}
\end{align}
where in~(\ref{Product1}), we used~(\ref{Permop}), in~(\ref{Perminvar}),  used the permutation invariance of $n$-dimensional integrals, in~(\ref{Conseq1}),  chose $\sigma=\tau^{-1}$, and finally in~(\ref{Conseq3}), used~(\ref{PermFunc}) and that $\mu_j$ and $\Lam_j$ are inverse of each other. Using~\eqref{Conseq3} in~\eqref{Lastbut2nd} we then find~\eqref{Nestcomm 2}.

\section{Permutation identities (II)}\label{sec:Pf3}

In this section, we prove~\eqref{pou} which is the most nontrivial step in the proof of~\eqref{iu}. 
Let us first note the identity 
\begin{align}
\text{lim}_{\epsilon\rightarrow 0}\int_{-\infty}^\infty dt\, \frac{\text{sinh}(\pi t)}{\text{sinh}(\pi(t\pm i\epsilon))}h(t) = \int_{-\infty}^\infty dt \,  h(t)
\end{align}
where $h(t)$ is a regular function at $t=0$. In subsequent manipulations, we will abbreviate identities of this type by dropping the integral and the limit as 
\begin{align}
\frac{\text{sinh}(\pi t)}{\text{sinh}(\pi(t\pm i\epsilon))}=1\label{Distrid}
\end{align}
which should (hopefully) cause no confusion. We also remind the reader of the short-hand notation introduced in~(\ref{shorthand}). For instance, from~\eqref{poi2} we have
\be \label{hyt}
F_m^\ep \equiv F_{\ep_i} (t_1, t_2, \cdots t_m) = f_1 g_{1,2,\epsilon_1} g_{2,3,\epsilon_2} \cdots g_{m-1,m,\epsilon_{m-1}} f_m 
\ee
and 
\be\label{yhn}
F^\epsilon_m = \biggl(\frac{g_{m-1,m,\epsilon_{m-1}}f_m}{f_{m-1}}\biggr)F^\epsilon_{m-1} \ .
\ee

\subsection{Preparation}

Before proving~\eqref{pou} we first prove a lemma. 

\textbf{Lemma}: Consider the operator
\begin{align}\label{nnp}
I_m[O]& \equiv \text{lim}_{\epsilon_i \rightarrow 0}\int F_m^\ep O(t_1,t_2,...,t_m), 
\end{align}
where $O (t_1, \cdots t_m)$ is a general operator and $F^\ep_m$ is defined in~\eqref{hyt}. For all $m \geq 2$, we have
\begin{align} \label{ppu}
\sum_{l=2}^m \Lambda_l I_m &=- I_m+\frac{1}{4}\sum_{l=2}^{m-1}R_l^{(m)}[O]
\end{align}
where $R^{(m)}_q[O]$ is defined in Eq.(\ref{contact2}). We remind the reader that $\Lam_l$ only acts on the kernel function $F$.

\textbf{Proof} : We prove this by induction on $m$. Taking $m=2$, $\Lambda_2 = (12)$, we get
\begin{equation}
\Lambda_2 I_2 = -I_2
\end{equation}
which is trivially true by the antisymmetry of the function $g$.  So the correct base case is $m=3$. 
Performing a few relabelings of the $t_i$, we get
\begin{align} \label{iii}
\Lambda_2 I_3+\Lambda_3 I_3 = \text{lim}_{\epsilon_1\rightarrow 0}\int dt_2 \int dt_1 \text{lim}_{\epsilon_2\rightarrow 0}\int dt_3\biggl(F_{\epsilon_1,\epsilon_2}(t_2,t_3,t_1)+F_{\epsilon_1,\epsilon_2}(t_2,t_1,t_3)\biggr)O(t_1,t_2,t_3)
\end{align}
Using the identity
\begin{align}
&4\text{cosh}(\pi t_2)\text{sinh}(\pi(t_1-t_2))\text{sinh}(\pi(t_2-t_3))\nonumber\\
&=-\text{cosh}(\pi(t_1-t_2+t_3))-\text{cosh}(\pi(3t_2-t_1-t_3))+\text{cosh}(\pi(t_3+t_2-t_1))+\text{cosh}(\pi(t_1-t_3+t_2))
\end{align}
and some algebraic manipulations we can write the term in the parentheses of~\eqref{iii} as 
\begin{align}
&F_{\epsilon_1,\epsilon_2}(t_2,t_3,t_1)+F_{\epsilon_1,\epsilon_2}(t_2,t_1,t_3) = \cr
&-4\biggl(\frac{i}{4}\biggr)\frac{f(t_1)f(t_2)f(t_3)\text{cosh}(\pi t_2)\text{sinh}(\pi(t_1-t_2))\text{sinh}(\pi(t_2-t_3))}{|\text{sinh}(\pi(t_3-t_2+i\epsilon_1))|^2|\text{sinh}(\pi(t_1-t_2+i\epsilon_1))|^2}g_{\epsilon_2}(t_1-t_3)\text{sinh}(\pi(t_1-t_3))\cr %\label{Line1}\\
&+\frac{i}{4}\frac{f(t_1)f(t_2)f(t_3)g_{\epsilon_2}(t_1-t_3)}{|\text{sinh}(\pi(t_3-t_2+i\epsilon_1))|^2|\text{sinh}(\pi(t_1-t_2+i\epsilon_1))|^2}\text{sinh}(\pi(t_1-t_3)) \cr
&\biggl[\text{cosh}(\pi(t_1-t_2+t_3))(\text{cosh}(3\pi i\epsilon_1)-1)-\text{cosh}(\pi(t_3+t_2-t_1))(\text{cosh}(i\pi\epsilon_1)-1)  \cr 
%\label{Line2}\\
&-\text{cosh}(\pi(t_1-t_3+t_2))(\text{cosh}(i\pi\epsilon_1)-1)+\text{cosh}(\pi(3t_2-t_1-t_3))(\text{cosh}(i\pi\epsilon_1)-1)\biggr] \ .\label{Line3}
\end{align}

%In deriving Eq.(\ref{Line1}), we used the identity

Integrating against $O(t_1,t_2,t_3)$ and taking $\epsilon_i\rightarrow 0$, we get
\begin{align}
& \Lambda_2 I_3+\Lambda_3 I_3 
%& \text{lim}_{\epsilon_1\rightarrow 0}\int dt_2 \int dt_1 \text{lim}_{\epsilon_2\rightarrow 0}\int dt_3\biggl[F_{\epsilon_1,\epsilon_2}(t_2,t_3,t_1)+F_{\epsilon_1,\epsilon_2}(t_2,t_1,t_3)\biggr]O(t_1,t_2,t_3)\cr
= -\text{lim}_{\epsilon_1\rightarrow 0}\int F_{\epsilon_1,\epsilon_1}(t_1,t_2,t_3)O(t_1,t_2,t_3) \cr
& -\frac{1}{8}\text{lim}_{\epsilon_1\rightarrow 0}\int O(t_1,t_2,t_3)  \, \frac{f(t_1)f(t_2)f(t_3)}{|\text{sinh}(\pi(t_3-t_2+i\epsilon_1))|^2|\text{sinh}(\pi(t_1-t_2+i\epsilon_1))|^2}\cr
&\times\biggl[\text{cosh}(\pi(t_1-t_2+t_3))(\text{cosh}(3\pi i\epsilon_1)-1)-\text{cosh}(\pi(t_3+t_2-t_1))(\text{cosh}(i\pi\epsilon_1)-1)\cr
&-\text{cosh}(\pi(t_1-t_3+t_2))(\text{cosh}(i\pi\epsilon_1)-1)+\text{cosh}(\pi(3t_2-t_1-t_3))(\text{cosh}(i\pi\epsilon_1)-1)\biggr]
\label{hgk}
\end{align}
The second term of~\eqref{hgk} can be simplified by noting the identity
\begin{align}
\text{lim}_{\epsilon\rightarrow 0}\int_{-\infty}^\infty dx \frac{\epsilon}{\epsilon^2+(x-a)^2}f(x) = \pi f(a)\ .
\end{align}
for $f$ continuous in $a$, bounded and integrable.
Thus, we obtain
\begin{align}
 \Lambda_2 I_3+\Lambda_3 I_3 
%& \text{lim}_{\epsilon_1\rightarrow 0}\int dt_2 \int dt_1 \text{lim}_{\epsilon_2\rightarrow 0}\int dt_3\biggl[F_{\epsilon_1,\epsilon_2}(t_2,t_3,t_1)+F_{\epsilon_1,\epsilon_2}(t_2,t_1,t_3)\biggr]O(t_1,t_2,t_3)\nonumber\\
= -\text{lim}_{\epsilon_1\rightarrow 0}\int F_{\epsilon_1,\epsilon_1}(t_1,t_2,t_3)O(t_1,t_2,t_3)+\frac{1}{4}\int_{-\infty}^\infty f(t)^2O(t,t,t) dt \ .
\end{align}
From the definitions (\ref{contact1})--(\ref{contact2}) we note that 
\begin{align}
R_2^{(3)}[O(t_1,t_2,t_3)] = \int dt f(t)^2 O(t,t,t)
\end{align}
and thus
\begin{align}
\sum_{j=2}^3\Lambda_j I_3[O] = -I_3[O]+\frac{1}{4}R_2^{(3)}[O], 
\end{align}
which establishes the lemma for the case $m=3$.

Suppose the lemma holds for some $m\geq 4$, i.e,
\begin{align}
\sum_{l=2}^{m-1}\Lambda_l I_{m-1}[O]=-I_{m-1}[O]+\frac{1}{4}\sum_{l=2}^{m-2}R_l^{(m-1)}[O]
\end{align}
Then, we have
\begin{align}
& \sum_{l=2}^m\Lambda_l I_{m}[O] = \text{lim}_{\epsilon\rightarrow 0}\sum_{j=2}^{m-2}\int\Lambda_l F_{m-1}^\epsilon\biggl(\frac{g_{m-1,m,\epsilon_{m-1}}f_m}{f_{m-1}}\biggr)O(t_1,...,t_m)\nonumber\\
&+\text{lim}_{\epsilon\rightarrow 0}\int \Lambda_m F_m^\epsilon O(t_1,..,t_m)+\text{lim}_{\epsilon\rightarrow 0}\int\frac{g_{1,m,\epsilon_{m-1}}f_m}{f_1}\Lambda_{m-1}F_{m-1}^\epsilon O(t_1,t_2,..,t_m) \ .
\end{align}
Rearranging the terms above gives
\begin{align}
& \sum_{l=2}^m\Lambda_l I_{m}[O] = \text{lim}_{\epsilon\rightarrow 0}\sum_{l=2}^{m-1}\int\Lambda_l F_{m-1}^\epsilon\biggl(\frac{g_{m-1,m,\epsilon_{m-1}}f_m}{f_{m-1}}\biggr)O(t_1,...,t_m)  +\text{lim}_{\epsilon\rightarrow 0}\int \Lambda_m F_m^\epsilon O(t_1,..,t_m)\cr
&+\text{lim}_{\epsilon_{m-1}\rightarrow 0}\int dt_m\text{lim}_{\epsilon\rightarrow 0}\int f_m\biggl(\frac{g_{1,m,\epsilon_{m-1}}}{f_1}-\frac{g_{m-1,m,\epsilon_{m-1}}}{f_{m-1}}\biggr)\Lambda_{m-1}F_{m-1}^\epsilon O(t_1,t_2,..,t_m) \ .
\label{428}
\end{align}
With a computation very similar  to the case $m=3$ we find the relation 
\begin{align}
&\text{lim}_{\epsilon\rightarrow 0}\int \Lambda_m F_m^\epsilon O(t_1,..,t_m)\\&+\text{lim}_{\epsilon_{m-1}\rightarrow 0}\int dt_m\text{lim}_{\epsilon\rightarrow 0}\int f_m\biggl(\frac{g_{1,m,\epsilon_{m-1}}}{f_1}-\frac{g_{m-1,m,\epsilon_{m-1}}}{f_{m-1}}\biggr)\Lambda_{m-1}F_{m-1}^\epsilon O(t_1,t_2,..,t_m) \\&= \frac{1}{4}R_{m-1}^{(m)}[O] \ . \label{SS}
\end{align}
Finally, we perform the integration over $t_m$ in the first term on right hand side of (\ref{428}) and use the induction hypothesis to obtain
\begin{align}
\sum_{l=2}^m\Lambda_l I_m[O]&=-I_m[O]+\frac{1}{4}\sum_{l=2}^{m-1}R_l^{(m)}[O]
\end{align}
which proves the lemma.

\subsection{Final proof}

We now prove~\eqref{pou} which using the definition~\eqref{nnp} can be written as 
\begin{align}
\Xi^m_k I_m[O]&=(-1)^k I_m[O]+\sum_{j=k+2}^m N_j[O]
\end{align}
with $N_j[O]$ given in Eq.(\ref{Ans1}).

Our strategy is to use induction on $m$ with a fixed $k$. First, consider $m = k+1$ for which we have 
\begin{align}
\Xi_k^{k+1} = \Lambda_m\Lambda_{m-1}\Lambda_{m-2}...\Lambda_2  \ .
\end{align}
It can be checked by an explicit computation
\begin{align}
\Xi_k^{k+1}  F_m = F(t_m,t_{m-1},t_{m-2},..,t_3,t_2,t_1)=(-1)^{m-1}F_m
\end{align}
following from $g_\ep (-t) = - g_\ep (t)$. This completes the proof for the case $m=k+1$.
Next, consider $m=k+2$, which is the base case for our induction argument.
By explicit computation one can show that 
\begin{align}
\Xi_k^{k+2} I_m [O] = \sum_{i_1=m-1}^m\sum_{i_2=m-2}^{i_1-1}....\sum_{i_{m-2}=2}^{i_{m-3}-1}\Lambda_{i_1}...\Lambda_{i_{m-2}}I_m[O]&=\sum_{l=2}^m(\Lambda_l I_m)[O']
\label{Basecase}
\end{align}
where $O'$ is defined as 
\begin{align}
%I_m[O'] &= \text{lim}_{\epsilon\rightarrow 0}\int F^\epsilon_m(y_1,...,y_m)O'(y_1,...y_m)\\
%t_k&=y_{m-k+1}\hspace{20pt} 1\leq k\leq m-1\\
O'(y_1,...y_m) \equiv O(y_m,y_{m-1},...,y_1) \ .
\end{align}
Now applying Lemma 1~\eqref{ppu} to the right hand side of~\eqref{Basecase} we find
\begin{align}
\Xi_k^{k+2} I_m [O] =-I_m[O']+\frac{1}{4}\sum_{j=1}^{k+1}R_j[O'] =(-1)^{k}I_m[O]+\frac{1}{4}\sum_{l=2}^{k+1}R_l^{(k+2)}[O']
\end{align}
where in the last step we have used the fact that $g_\ep (t)$ is an odd function. This completes the proof for 
$m=k+2$. 
 
Now, suppose~\eqref{pou} holds for $m \geq k+3$. Remember the definition of $\Xi^m_k$ from ~\eqref{bbu}. It can be checked explicitly that $\Xi^m_k = \Xi_k^{m-1} + \Lam_m \Xi_{k-1}^{m-1}$. 
Then, we have
\begin{align}
\Xi_k^m F_m^\ep & = \sum_{i_1=k+1}^{m}\sum_{i_2=k}^{i_1-1}\sum_{i_3=k-1}^{i_2-1}...\sum_{i_k=2}^{i_{k-1}-1}\Lambda_{i_1}\Lambda_{i_2}...\Lambda_{i_k} F^\epsilon_{m}\\
&=\Xi_k^{m-2} \biggl(F^\epsilon_{m-1}\frac{g_{m-1,m}f_m}{f_{m-1}}\biggr) + \Lam_{m-1} \Xi_{k-1}^{m-2} \biggl(F^\epsilon_{m-1}\frac{g_{m-1,m}f_m}{f_{m-1}}\biggr) 
+\Lambda_{m} \Xi_{k-1}^{m-1} F^\epsilon_m 
\\&=\frac{g_{m-1,m}f_m}{f_{m-1}}\Xi_k^{m-1} F^\epsilon_{m-1}+\Lambda_{m} \Xi_{k-1}^{m-1} F^\epsilon_m +C \Lambda_{m-1}\Xi_{k-1}^{m-2} F^\epsilon_{m-1} \
%&=\frac{g_{m-1,m}f_m}{f_{m-1}}\sum_{i_1=k+1}^{m-1}\sum_{i_2,..,i_k}\Lambda_{i_1}\Lambda_{i_2}..\Lambda_{i_k}F^\epsilon_{m-1}+\sum_{p=1}^{k-1} S_p+R_k
\label{nuj}
\end{align}
%where we have used the definition~\eqref{bbu}.

Here,
\begin{align}
C&=\frac{g_{1,m}f_m}{f_{1}}-\frac{g_{m-1,m}f_m}{f_{m-1}}\label{AA}\ .
\end{align}
If we integrate against $O(t_1,t_2,..,t_m)$ and take $\epsilon_i \rightarrow 0$, the first term in~\eqref{nuj} goes to 
\begin{align}
(-1)^k I_m[O] +\sum_{j=k+2}^{m-1} N_j[O]
\end{align}
by the induction hypothesis. To finish the proof we need to show that the second and the third term add up to
$N_m[O]$.

%Note that this is not as straightforward as the derivation of the relation  in~\eqref{SS}. 

Let us denote the sum of the second and the third terms in~\eqref{nuj} as $V$.
%The function $C$ in ~\eqref{AA} contains a factor $\sinh(\pi(t_{m-1}-t_1))$, but the second term of~\eqref{nuj} does not contain such an overall factor. 
%So the story is a bit intricate, and we will do it in several steps. 
First, split $V$ into 
\begin{align}
V&=S_1+V_1\\
S_1&= \Lambda_m \Xi^{m-3}_{k-1}F_m^\epsilon+C\Lambda_{m-1}\Xi^{m-3}_{k-1}F_{m-1}^\epsilon\\
V_1&=\Lambda_m\Lambda_{m-1}\Xi_{k-2}^{m-2}F_m^\epsilon+\Lambda_m\Lambda_{m-2}\Xi_{k-2}^{m-3}F_m^\epsilon+C \Lambda_{m-1}\Lambda_{m-2}\Xi_{k-2}^{m-3}F_{m-1}^\epsilon
\end{align}
where we have repeatedly used $\Xi^m_k = \Lambda_m \Xi^{m-1}_{k-1}+\Xi^{m-1}_k$.
%{\label{Definition}
%\end{align}}
Let us rename $y_1=t_m$, $y_2=t_{m-1}$ and $y_3=t_1$ for the moment.
We find
\begin{align}
S_1&=\frac{\chi_{1m}(t)'}{f_{m-1}}\sum_{l=1}^3\Lambda_l F_3^\epsilon(y_1,y_2,y_3)\\
\chi_{1m}(t)'&=\Lambda_{m-1}\sum_{i_2=k}^{m-3}\sum_{i_3=k-1}^{i_2-1}..\sum_{i_k=2}^{i_{k-1}-1}\Lambda_{i_2}..\Lambda_{i_k}(f_1 g_{1,2,\epsilon_1}g_{2,3,\epsilon_2}...g_{m-3,m-2,\epsilon_{m-3}})\ .
\end{align}

Integrating $S_1$ against $O(t_1,...,t_m)$ and taking $\epsilon\rightarrow 0$, we find this is precisely the $p=1,j=m$ term in $N_j[O]$ that we are looking for, after applying the Lemma, in Eq.(\ref{Ans1}).

Now, $V_1$ can be further split into
\begin{align}
V_1 &= S_2+V_2\\
S_2 &= \Lambda_m\Theta^m_2 \Xi^{m-4}_{k-2}F_m^\epsilon+C \Lambda_{m-1}\Lambda_{m-2}\Xi^{m-4}_{k-2}F^\epsilon_{m-1}\label{S}\\
V_2&=\Lambda_m\Theta^m_3\Xi^{m-4}_{k-3}F^\epsilon_m+\Lambda_m\Lambda_{m-1}\Lambda_{m-2}\Lambda_{m-3}\Xi^{m-4}_{k-4}F^\epsilon_m +C\Lambda_{m-1}\Lambda_{m-2}\Lambda_{m-3}\Xi^{m-4}_{k-3}F^\epsilon_{m-1}
\end{align} 
where
\bln
\Theta^m_2 & \equiv  \Lambda_{m-1}+\Lambda_{m-2}\\
\Theta^m_3 & \equiv \Lambda_{m-1}\Lambda_{m-2}+\Lambda_{m-1}\Lambda_{m-3}+\Lambda_{m-2}\Lambda_{m-3} = \sum_{i_1=m-2}^{m-1}\sum_{i_2=m-3}^{i_1-1}\Lambda_{i_1}\Lambda_{i_2} \ .
\label{Q}
\end{align}
It is convenient to rename $y_1=t_m,$ $y_2=t_{m-1}$, $y_3=t_2$ and $y_4=t_1$. Now, using the equalities
\begin{align}
&\Lambda_m\Lambda_{m-1}\Xi^{m-4}_{k-1}F_m^\epsilon=\chi_{2m}' g_{m,m-1}g_{2,m}g_{1,2}f_1\\
&\Lambda_m\Lambda_{m-2}\Xi^{m-4}_{k-1}F_m^\epsilon=\chi_{2m}' g_{2,m-1}g_{m,2}g_{1,m}f_1\\
&\Lambda_{m-1}\Lambda_{m-2}\Xi^{m-4}_{k-1}F^\epsilon_{m-1}=\chi_{2m}' g_{2,m-1}g_{1,2}f_1
\end{align}
we can write $S_2$ as 
\be \label{S2eq}
S_2 = \frac{\chi_{2m}'}{f_{m-1}} \le(\sum_{j=1}^4\Lambda_j F_4 (y_1,y_2,y_3,y_4) \ri)\ .
\ee
%where in the first equality  and in the second equality used the Lemma. 
Note that since 
\begin{align}
\chi_{2m}'(t)&=\Lambda_{m-1}\Lambda_{m-2}\sum_{i_3=k-1}^{m-4}\sum_{i_4=k-2}^{i_2-1}..\sum_{i_k=2}^{i_{k-1}-1}\Lambda_{i_2}..\Lambda_{i_k}(f_1 g_{1,2,\epsilon_1}g_{2,3,\epsilon_2}...g_{m-4,m-3,\epsilon_{m-4}})
\end{align}
if we integrate $S_2$ in (\ref{S2eq}) against $O(t_1,..,t_m)$ and take all the $\epsilon\rightarrow 0$ we find the $p=2, j=m$ term in $N_j[O]$ in Eq.(\ref{Ans1}). 

Now, the pattern is clear:  we split the remainder $V_j$ into $S_{j+1}$ and $V_{j+1}$ until $j=k$. Since $V_k=0$ we have
\begin{align} \label{mnb}
V=\sum_{p=1}^{k}S_p  , 
\quad
S_p= \Lambda_m\Theta^m_p\Xi^{m-p-2}_{k-p}F^\epsilon_m +C\Lambda_{m-1}\Lambda_{m-2}\cdots \Lambda_{m-p}\Xi^{m-p-2}_{k-p}F^\epsilon_{m-1}, %\quad 1 \leq p \leq k-1 
%
%V_{k-1}&=\Lambda_m\Theta^m_k F^\epsilon_m+B\Lambda_{m-1}\Lambda_{m-2}..\Lambda_{m-k}F^\epsilon_{m-1}
\end{align}
where 
\begin{align}
\Theta^m_p=\sum_{i_1=m-2}^{m-1}\sum_{i_2=m-3}^{i_1-1}\cdots\sum_{i_{p-1}=m-p}^{i_{p-2}-1}\Lambda_{i_1}\cdots \Lambda_{i_{p-1}} \ .
\end{align}
Note that $\Theta^m_p$ is a direct generalization of~(\ref{Q})) and $\Xi_0^j=1$. 

The basic idea to compute $S_p$ is the same as that of $S_2$. One has to show that $S_p$ can be written as 
\be \label{hine}
S_p = \frac{\chi_{pm}'}{f_{m-1}}\le(\sum_{j=1}^{p+2}\Lambda_j F_{p+2}(y_1, y_2, \cdots y_{p+2}) \ri)
\ee
with 
\begin{align}\label{yvar}
y_1=t_m, \quad  y_2= t_{m-1}, \quad y_{3+q} = t_{p-q} \;\; {\rm for} \;\;  0\leq q\leq p-1 \ .
\end{align} 
It is important to remember that $\chi'_{pm}$ does not depend on any of the $y$ variables. It then follows from the Lemma that after integrating against $O(t_1,..,t_m)$ and sending $\epsilon\rightarrow 0$, $S_p $ has the form required in Eq.(\ref{Ans1}). 

To finish the proof, we need to demonstrate~\eqref{hine}.\footnote{It is enough to take $k \geq 3$ as the previous computations of $S_{1,2}$ are sufficent to cover $k=2$.} 
For this purpose, let us look at the first term of $S_p$ in~\eqref{mnb}:
\be \label{s11}
\Lambda_m\Theta^m_p\Xi^{m-p-2}_{k-p}F^\epsilon_m  \ .
\ee
Expanding $\Xi$ explicitly we have 
\bea \label{Perm10}
&&\Xi^{m-p-2}_{k-p}F_m^\epsilon = \sum_{i_{p+1}=k-p+1}^{m-p-2}\sum_{i_{p+2}=k-p}^{i_{p+1}-1}...\sum_{i_k=2}^{i_{k-1}-1}\Lambda_{i_{p+1}}..\Lambda_{i_k}F_m^\epsilon 
\eea
and
\bea
%\begin{align}
&&\Lambda_{i_{p+1}}\Lambda_{i_{p+2}}....\Lambda_{i_k} F^\epsilon_m = \chi_1g_{m-p-1,m-p}...g_{m-1,m}f_m\label{Perm1a}\\
%\Lambda_{i_{p+1}}\Lambda_{i_{p+2}}....\Lambda_{i_k} F^\epsilon_{m-1}&= \chi_1g_{m-p-1,m-p}...g_{m-2,m-1}f_{m-1}\label{Perm1b}\\
&&\chi_1=f_{\tau(1)}g_{\tau(1)\tau(2)}g_{\tau(2)\tau(3)}...g_{\tau(m-p-3),\tau(m-p-2)}g_{\tau(m-p-2),m-p-1}
\eea
where the permutation $\tau$ and the function $\chi_1$ depend on the multi-index $i_{p+1},\cdots,i_k$. It is crucial that the permutation $\tau$ does not touch any of the indices above $m-p-2$. To lighten the notation, we suppress the $i$ dependence of $\chi_1$. Equation~\eqref{Perm10}--\eqref{Perm1a} further imply 
\begin{align} 
\Xi^{m-p-2}_{k-p}F^\epsilon_m&=\chi_2g_{m-p-1,m-p}...g_{m-1,m}f_m\label{conc1}
%\Xi^{m-p-2}_{k-p}F^\epsilon_{m-1}&=\chi_2g_{m-p-1,m-p}...g_{m-2,m-1}f_{m-1} \label{conc2}
\end{align}
where we have introduced
\begin{align}
\chi_2&=\sum_{i_{p+1}=k-p+1}^{m-p-2}\sum_{i_{p+2}=k-p}^{i_{p+1}-1}...\sum_{i_k=2}^{i_{k-1}-1}\chi_1 \ .
\end{align}
Next, note that
\begin{align} \label{Perm2}
\Lambda_m\Theta^m_p &= \Lambda_m\sum_{j=0}^{p-1}K_j, \quad 
K_j =\biggl(\Pi_{l=1}^j\Lambda_{m-l}\biggr)\biggl(\Pi_{n=j+1}^{p-1}\Lambda_{m-n-1}\biggr) \ .
\end{align} 
Every cyclic permutation appearing in~(\ref{Perm2}), has the exact same action on $\chi_2$, therefore
\be 
\Lambda_mK_j\chi_2 =\chi_{pm}'= \Lambda_{m-1}..\Lambda_{m-p}\chi_2 \ .
\label{Def}
\ee
The first equality of~\eqref{Def} says that the action of $\Lam_m K_j $ on $\chi_2$ is independent of $j$ and we have named the resulting function $\chi_{pm}'$ in (\ref{hine}). The second equality says $\chi_{pm}'$ can also be obtained from the permutation $ \Lambda_{m-1}\cdots\Lambda_{m-p}$ acting on $\chi_2$. The reasons behind~\eqref{Def} are: (i)  the number of cycles in each permutation string acting on $\chi_2$ is the same; (ii) the length of each cycle is larger than the highest index occurring in $\chi_2$. As a simple example, consider the following permutation on some function $G(t_1,t_2,t_3)$
\be 
\Lambda_{10}\Lambda_9\Lambda_8 G(t_1,t_2,t_3) = \Lambda_6\Lambda_5\Lambda_4 G(t_1,t_2,t_3) \ .
\ee

The $\chi_{pm}'$ in Eq.(\ref{Def}) has the precise expression we saw for the $j=m$ contact contribution in Eq.(\ref{Ans1}) which is
\be
\chi'_{pm} = \Lambda_{m-1}\cdots\Lambda_{m-p}\sum_{i_{p+1}=k-p+1}^{m-p-2}\sum_{i_{p+2}=k-p}^{i_{p+1}-1}\cdots\sum_{i_k=2}^{i_{k-1}-1}\Lambda_{i_{p+1}}\Lambda_{i_{p+2}}....\Lambda_{i_k}(f_1g_{1,2,\epsilon_1}g_{2,3,\epsilon_2}...g_{m-p-2,m-p-1,\epsilon_{m-p-2}})
\ee
Now, let us look at how $K_j$ act on the product of the functions in~\eqref{conc1} other than $\chi_2$. 
To this end, we define 
\begin{align}
G (u_1,u_2,..,u_l) \equiv g_{\epsilon_1}(u_2-u_1)g_{\epsilon_2}(u_3-u_2)...g_{\epsilon_{l-1}}(u_l-u_{l-1}) \ .
\end{align}
We then find 
\be
\Lambda_m\sum_{j=0}^{p-1}K_j G(t_{m-p-1},t_{m-p},...t_{m-1},t_m)=\sum_{j=0}^{p-1}G(t_{m-1},t_{p},t_{p-1}...,t_{j+2},t_{m},t_{j+1},t_j,...,t_2,t_1) \ .
\ee
%and 
%\begin{gather}
%\Lam_m K_{0} G(t_{m-p-1},t_{m-p},...,t_{m-1},t_m)=
%G (t_{m-1},t_p,t_{p-1},..,t_2,t_m,t_1)\\
%\Lam_m K_{p-1} G(t_{m-p-1},t_{m-p},...,t_{m-1},t_m)=G(t_{m-1},t_m,t_p,t_{p-1},...,t_2,t_1) \ .
%\end{gather}
Collecting everything together we thus find~\eqref{s11} can be written as 
\bega \label{ff1}
\Lambda_m\Theta^m_p\Xi^{m-p-2}_{k-p}F^\epsilon_m  =\chi_{pm}' f_1 \sum_{j=0}^{p-1} G(t_{m-1},t_p,t_{p-1},...,t_{j+2},t_m,t_{j+1},t_j...,t_2,t_1)  \ . %\ri. \cr
%+ G (t_{m-1},t_p,t_{p-1},...,t_2,t_m,t_1) +H(t_{m-1},t_m,t_p,t_{p-1}...,t_2,t_1)
\end{gather}

The second term of $S_p$ in~\eqref{mnb} can be treated exactly in parallel, and we find 
\be\label{ff2} 
C\Lambda_{m-1}\Lambda_{m-2}..\Lambda_{m-p}\Xi^{m-p-2}_{k-p}F^\epsilon_{m-1} =\chi_{pm}' f_1 C 
G(t_{m-1},t_p,t_{p-1},..,t_2,t_1)
\ee
Plugging the explicit expression for $C$, combining~\eqref{ff1} and~\eqref{ff2}, cancelling some terms, and relabeling 
the arguments as in~\eqref{yvar}, we then obtain~\eqref{hine}. This completes the proof.

\vspace{0.2in}   \centerline{\bf{Acknowledgements}} \vspace{0.2in}
We thank Tom Faulkner and Daniel Harlow for discussions. SR would like to thank his undergrad mathematics instructor Prof. Subiman Kundu of IIT Delhi for his great lectures on functional analysis.
This work is partially supported by the Office of High Energy Physics of U.S. Department of Energy under grant Contract Numbers DE-SC0012567 and DE-SC0019127. The work of NL is supported by a grant-in-aid from the National Science Foundation grant 
number PHY-1606531.

\appendix

\section{Harmonic oscillator example} \label{app:b}

As an illustration of the perturbation series~\eqref{nju} here we consider a very simple example. 

Consider $\De_0$ given by the (unnormalized) thermal density matrix  for a harmonic oscillator
\begin{align}
\Delta_0&= \text{exp}(-\beta H), \qquad 
H=\hat{a}^\dagger \hat{a}
\end{align}
and $\De$ by a ``squeezed transform'' of $\De_0$ 
\begin{align}
\Delta =S(r)^\dagger\text{exp}(-\beta H)S(r), \quad 
S(r) = \text{exp}\biggl(\frac{1}{2}r\hat{a}^2-\frac{1}{2}r\hat{a}^{\dagger 2}\biggr), \hspace{20pt}r>0 \ .
\end{align}

We will compute $\text{log}\Delta_0-\text{log}\Delta$ as a perturbative expansion in the parameter $r$ two ways, firstly using the
the BCH (which in this example is extremely simple) and then using the formula~\eqref{intrep0}.

From BCH, it is straightforward to see that
\begin{align}
\Delta =\text{exp}(-\beta H'), \quad 
H' = S^\dagger H S  =\sigma^\dagger\sigma, \quad 
\sigma =S^\dagger\hat{a}S = \hat{a}\text{cosh}r-\hat{a}^\dagger\text{sinh}r , 
\end{align}
and thus 
\be
\text{log}\Delta_0-\text{log}\Delta = \beta(H'-H) 
= (2\hat{a}^\dagger\hat{a}+1)\biggl(\beta r^2+\frac{\beta r^4}{3}\biggr)-(\hat{a}^2+\hat{a}^{\dagger 2})\biggl(r\beta+\frac{2\beta r^3}{3}\biggr)+\mathcal{O}(r^5) \ .
\label{Series}
\ee

Now let us consider the perturbation series~\eqref{intrep0}. Note that 
\begin{align}
\Delta_0^{-it}\hat{a}\Delta_0^{it} =\hat{a}e^{-i\beta t}, \quad 
\Delta_0^{-1/2}\hat{a}\Delta_0^{1/2}=\hat{a}e^{-\beta/2}, \quad \Delta_0^{1/2}\hat{a}\Delta_0^{-1/2}=\hat{a}e^{\beta/2}
\end{align}
Here,  the domain one should consider is the set of finitely excited Harmonic oscillator states. Throughout, we will be working on this set. Then, we get
\begin{align}
\bm{\alpha} &= 1-\Delta_0^{-1/2}\Delta_1\Delta_0^{-1/2}\\&=\frac{1}{2}r\Delta_0^{-1/2}[\hat{a}^2-\hat{a}^{\dagger 2},\Delta_0]\Delta_0^{-1/2}-\frac{1}{8}r^2\Delta_0^{-1/2}[\hat{a}^2-\hat{a}^{\dagger 2},[\hat{a}^2-\hat{a}^{\dagger 2},\Delta_0]]\Delta_0^{-1/2}\\&+\frac{r^3}{3!2^3}\Delta_0^{-1/2}[\hat{a}^2-\hat{a}^{\dagger 2},[\hat{a}^2-\hat{a}^{\dagger 2},[\hat{a}^2-\hat{a}^{\dagger 2},\Delta_0]]\Delta_0^{-1/2}\\&-\frac{r^4}{2^44!}\Delta_0^{-1/2}[\hat{a}^2-\hat{a}^{\dagger 2},[\hat{a}^2-\hat{a}^{\dagger 2},[\hat{a}^2-\hat{a}^{\dagger 2},[\hat{a}^2-\hat{a}^{\dagger 2},\Delta_0]\Delta_0^{-1/2},\\
\bm{\delta}&=\frac{\bm{\alpha}}{1-\bm{\alpha}/2}\\
&=-r\text{sinh}\beta(\hat{a}^2+\hat{a}^{\dagger 2})+\frac{r^2}{4}[\hat{a}^2,\hat{a}^{\dagger 2}]\text{sinh}(2\beta)+\frac{r^3}{12}\text{sinh}^3\beta(\hat{a}^2+\hat{a}^{\dagger 2})^3\\&-\frac{r^3}{24}\text{cosh}\beta\text{sinh}(2\beta)[\hat{a}^2-\hat{a}^{\dagger 2},[\hat{a}^2,\hat{a}^{\dagger 2}]]\\&-\frac{r^4}{3.2^4}\text{sinh}^2\beta\text{sinh}2\beta\biggl((\hat{a}^2+\hat{a}^{\dagger 2})^2[\hat{a}^2,\hat{a}^{\dagger 2}]+[\hat{a}^2,\hat{a}^{\dagger 2}](\hat{a}^2+\hat{a}^{\dagger 2})^2+(\hat{a}^2+\hat{a}^{\dagger 2})[\hat{a}^2,\hat{a}^{\dagger 2}](\hat{a}^2+\hat{a}^{\dagger 2})\biggr)\\&+\text{sinh}2\beta\frac{r^4}{4!2^3}\biggl(\text{cosh}^2\beta[[[\hat{a}^2,\hat{a}^{\dagger 2}],\hat{a}^2-\hat{a}^{\dagger 2}],\hat{a}^2-\hat{a}^{\dagger 2}]-\text{sinh}^2\beta[[[\hat{a}^2,\hat{a}^{\dagger 2}],\hat{a}^2+\hat{a}^{\dagger 2}],\hat{a}^2+\hat{a}^{\dagger 2}]\biggr).
\end{align}

We define
\begin{align}
s(t)&= \hat{a}^2 e^{-2i\beta t}+\hat{a}^{\dagger 2}e^{2i\beta t},\hspace{20pt}d(t)=\hat{a}^2 e^{-2i\beta t}-\hat{a}^{\dagger 2}e^{2i\beta t}\\
\end{align}
in terms of which we obtain
\begin{align}
\bm{\delta}(t)&=-r\text{sinh}\beta s(t)+\frac{r^2}{4}[\hat{a}^2,\hat{a}^{\dagger 2}]\text{sinh}2\beta\\&+\frac{r^3}{12}\text{sinh}^3\beta s(t)^3-\frac{r^3}{24}\text{cosh}\beta\text{sinh}(2\beta)[d(t),[\hat{a}^2,\hat{a}^{\dagger 2}]]\\&-\frac{r^4}{3.2^4}\text{sinh}^2\beta\text{sinh}2\beta\biggl(s(t)^2[\hat{a}^2,\hat{a}^{\dagger 2}]+[\hat{a}^2,\hat{a}^{\dagger 2}]s(t)^2+s(t)[\hat{a}^2,\hat{a}^{\dagger 2}]s(t)\biggr)\\&+\text{sinh}2\beta\frac{r^4}{4!2^3}\biggl(\text{cosh}^2\beta[[[\hat{a}^2,\hat{a}^{\dagger 2}],d(t)],d(t)]-\text{sinh}^2\beta[[[\hat{a}^2,\hat{a}^{\dagger 2}],s(t)],s(t)]\biggr).
\end{align}
We would need the expressions
\begin{align}
\int dt\frac{e^{ixt}}{\text{cosh}^2(\pi t)}&=\frac{x}{\pi}\frac{1}{\text{sinh}(x/2)},\\
\text{lim}_{\epsilon\rightarrow 0}\int\frac{dt}{\text{cosh}(\pi t)}\frac{e^{ix(t-a)}}{\text{sinh}(\pi(t-a+i\epsilon))}&=2i\frac{e^{-x/2}-e^{-iax}}{1-e^{-x}}\frac{e^{-x/2}}{\text{cosh}(\pi a)},\\
\text{lim}_{\epsilon\rightarrow 0}\int\frac{dt}{\text{cosh}(\pi t)}\frac{e^{ix(t-a)}}{\text{sinh}(\pi(t-a-i\epsilon))}&=2i\frac{e^{x/2}-e^{-iax}}{1-e^{-x}}\frac{e^{-x/2}}{\text{cosh}(\pi a)}.
\end{align}

Plugging all this in the expressions in (\ref{firsteq})--(\ref{fourtheq}), and adding all the contributions, we get, upto $\mathcal{O}(r^4)$,
\begin{align}
\sum_{i}Q_i&=-r\beta(\hat{a}^2+\hat{a}^{\dagger 2})+\frac{1}{2}\beta r^2[\hat{a}^2,\hat{a}^{\dagger 2}]-\frac{\beta r^3}{12}[\hat{a}^2-\hat{a}^{\dagger 2},[\hat{a}^2,\hat{a}^{\dagger 2}]]\\&-\frac{\beta r^4}{96}([[[\hat{a}^2,\hat{a}^{\dagger 2}],\hat{a}^2],\hat{a}^{\dagger 2}]+[[[\hat{a}^2,\hat{a}^{\dagger 2}],\hat{a}^{ \dagger 2}],\hat{a}^{2}])\\&=(2\hat{a}^\dagger\hat{a}+1)\biggl(\beta r^2+\frac{\beta r^4}{3}\biggr)-(\hat{a}^2+\hat{a}^{\dagger 2})\biggl(r\beta+\frac{2\beta r^3}{3}\biggr).\label{Answer}
\end{align}
where we have combined terms using the identity
\begin{align}
[s_1,s_2^3]-3s_2[s_1,s_2]s_2=[[[s_1,s_2],s_2],s_2] \ .
\end{align}
Eq.(\ref{Answer}) and Eq.(\ref{Series}) agree precisely.

\section{Contact term at quintic order}\label{sec:quintic}

For completeness, we evaluate the formula given for the contact terms $P_m$ for the case $m=5$. We get
\begin{align}
P_5&=\frac{\pi}{40}\text{lim}_{\epsilon_1,\epsilon_2\rightarrow 0}\int \frac{g_{\epsilon_1}(t_2-t_1)g_{\epsilon_2}(t_3-t_2)}{\text{cosh}(\pi t_1)\text{cosh}(\pi t_3)}\bigg(\bm{\delta}(t_1)^3\bm{\delta}(t_2)\bm{\delta}(t_3)+\bm{\delta}(t_2)\bm{\delta}(t_1)\bm{\delta}(t_2)^2\bm{\delta}(t_3)\nonumber\\
&+\bm{\delta}(t_1)\bm{\delta}(t_3)\bm{\delta}(t_2)\bm{\delta}(t_1)^2+\bm{\delta}(t_1)\bm{\delta}(t_2)\bm{\delta}(t_3)^3+\bm{\delta}(t_1)\bm{\delta}(t_2)^2\bm{\delta}(t_3)\bm{\delta}(t_2)\nonumber\\
&+\bm{\delta}(t_1)^2\bm{\delta}(t_2)\bm{\delta}(t_3)\bm{\delta}(t_1)-\bm{\delta}(t_1)\bm{\delta}(t_2)\bm{\delta}(t_3)\bm{\delta}(t_1)^2-\bm{\delta}(t_2)^2\bm{\delta}(t_3)\bm{\delta}(t_2)\bm{\delta}(t_1)\nonumber\\
&-\bm{\delta}(t_1)^2\bm{\delta}(t_3)\bm{\delta}(t_2)\bm{\delta}(t_1)-\bm{\delta}(t_1)\bm{\delta}(t_2)\bm{\delta}(t_3)\bm{\delta}(t_2)^2+\bm{\delta}(t_1)\bm{\delta}(t_2)^3\bm{\delta}(t_3)\biggr).
\end{align}
The first three terms above come from the object $\Xi^5_1$ applied to the kernel. The next three terms come from $\Xi^5_3$. The last five terms come from $\Xi^5_2$.

This can be further simplified as
\begin{align}
P_5&=\frac{\pi}{40}\text{lim}_{\epsilon_1,\epsilon_2\rightarrow 0}\int \frac{g_{\epsilon_1}(t_2-t_1)g_{\epsilon_2}(t_3-t_2)}{\text{cosh}(\pi t_1)\text{cosh}(\pi t_3)}\bigg([\bm{\delta}(t_2)\bm{\delta}(t_1),\bm{\delta}(t_2)^2\bm{\delta}(t_3)]\nonumber\\
&+\bm{\delta}(t_1)[\bm{\delta}(t_3),\bm{\delta}(t_2)]\bm{\delta}(t_1)^2+\bm{\delta}(t_1)\bm{\delta}(t_2)[\bm{\delta}(t_2),\bm{\delta}(t_3)]\bm{\delta}(t_2)+\bm{\delta}(t_1)^2[\bm{\delta}(t_2),\bm{\delta}(t_3)]\bm{\delta}(t_1)\nonumber\\
&+\bm{\delta}(t_1)\bm{\delta}(t_2)^3\bm{\delta}(t_3)+\bm{\delta}(t_1)^3\bm{\delta}(t_2)\bm{\delta}(t_3)+\bm{\delta}(t_1)\bm{\delta}(t_2)\bm{\delta}(t_3)^3\biggr).
\end{align}

\section{Action of the Permutation Group on Operators and Functions} \label{app:a}

In this Appendix, we elaborate more on the action of the permutation group 
on operators and functions introduced in~\eqref{i1}--\eqref{i2}.

$S_m$ denotes the symmetric group of permutations on $m$-distinct objects and we use $(1,2,..,m)$ to denote the permutation that sends $1\rightarrow 2\rightarrow 3 \cdots \to (m-1)\rightarrow m\rightarrow 1$. 
We will follow the convention where group composition, denoted by $\star$ in $S_m$ is from left to right, e.g., 
\begin{align}
(1234)\star(124) &= (1423)\\
(12345)\star(342)&=(13245)
\end{align}

We define 
\begin{align}\label{i3}
\sigma(\bde(t_{i_1})\bde(t_{i_2})\cdots \bde(t_{i_m}))&= \bde(t_{i_{\sigma (1)}})\bde(t_{i_{\sig(2)}})\cdots\bde(t_{i_{\sig(m)}})\\
\sigma(F(t_{i_1},t_{i_2},\cdots,t_{i_m}))&= F(t_{\sigma{(i_1)}},t_{\sigma(i_2)},\cdots,t_{\sigma(i_m)}) \ .
\label{i4}
\end{align}
Note that in~\eqref{i3}  the permutations act to the left, while
in~\eqref{i4}, the permutations act to the right. More explicitly, 
\bega
\sigma(\tau(\bde(t_{i_1})\bde(t_{i_2}) \cdots\bde(t_{i_m}))) = \sigma\star\tau\bigl(\bde(t_{i_1})\bde(t_{i_2})\cdots \bde(t_{i_m})\bigr) , %= \bde(t_{i_{\sigma(\tau(1))}})\bde(t_{i_{\sigma(\tau(2))}}) \cdots \bde(t_{i_{\sigma(\tau(m))}}),
\label{Permop}\\
\sigma(\tau(F(t_{i_1},t_{i_2},\cdots,t_{i_m})))=\tau\star\sigma\bigl(F(t_{i_1},t_{i_2},\cdots ,t_{i_m}))\bigr) %= F(t_{\sigma(\tau(i_1))}, t_{\sigma(\tau(i_2))},\cdots,t_{\sigma\tau((i_m)) }) 
\ .
\label{PermFunc}
\end{gather}

Now consider some examples 
\bega
(123)F(t_2,t_1,t_3)= F(t_3,t_2,t_1) , \quad (123)\bde(t_2)\bde(t_1)\bde(t_3)=\bde(t_1)\bde(t_3)\bde(t_2),   \\
(12) ((123) F(t_2,t_1,t_3)) = F(t_3,t_1,t_2) = (23) F(t_2,t_1,t_3) , \\
 (12) ((123)\bde(t_2)\bde(t_1)\bde(t_3))=\bde(t_3)\bde(t_1)\bde(t_2) = (13) \bde(t_2)\bde(t_1)\bde(t_3),
\end{gather}
with $(12)\star (123) = (13)$ and $(123)\star (12) = (23)$. 

We now give a proof Eq.~(\ref{Nestcomm 1})~\cite{Blessenohl:1988}. 
Notice that 
\begin{align}
(m,m-1...4321)X\bde(t_m)=\bde(t_m)X, 
\end{align}
for \textit{any} product of operators $X$ not involving $\bde(t_m)$, and 
thus 
\be 
 ({\rm id}-(m,m-1,..4321)) X\bde(t_m)
= [X, \bde (t_m)] \ .
\ee
Now~\eqref{Nestcomm 1} follows by induction.
 For $m=2$ it is obviously true. Assume its true for $m-1$. Then we have
\begin{align}
T_m(\bde(t_1)..\bde(t_m))&=({\rm id}-(m,m-1,..4321))\biggl[T_{m-1}(\bde(t_1)..\bde(t_{m-1}))\biggr]\bde(t_m) \cr
& = [T_{m-1}(\bde(t_1)..\bde(t_{m-1})), \bde(t_m)]
%\\&=(\text{id}-(m,m-1,..4321))\biggl([..[[\bde(t_1),\bde(t_2)],\bde(t_3)],...,],\bde(t_{m-1})]\biggr)\bde(t_m)
\end{align}
which completes the proof.

\section{The $\epsilon\rightarrow 0$ limit}\label{sec:Pf4}

In this appendix, we include a proof of the existence of the $\epsilon\to 0$ limit in (\ref{intrep0}). 
In other words, we establish the identity in (\ref{intrep3}) treating the $\epsilon\to 0$ limit carefully. 

Using the spectral decomposition of $\Delta$ in (\ref{spectral}) for any $\ep,\beta>0$ and vectors $|x\rangle$ and $|y\rangle$ we have
\begin{align}
\lim_{\ep\to 0}\:\langle x|\frac{\Delta^{1-\ep}}{\Delta+\beta}|y\rangle=\frac{-i}{2}\text{lim}_{\epsilon\rightarrow 0}\int_{-\infty}^\infty \frac{dt}{\text{sinh}(\pi(t-i\epsilon))}\beta^{-it}\langle y| \Delta^{-it}|x\rangle 
\end{align}
Note that we can freely interchange the $\lam$ and $t$ integrals because the integrand above is an absolutely convergent function. Our goal is to show the limit above gives $\langle y|\frac{\Delta}{\Delta+\beta}|x\rangle$. 

The following inequalities
\begin{align}
\frac{e^{(1-\epsilon)\lambda}}{e^\lambda+\beta}<\frac{e^\lambda}{e^\lambda+\beta}<1\hspace{20pt}\forall \lambda>0,\\
\frac{e^{(1-\epsilon)\lambda}}{e^\lambda+\beta}<\frac{1}{e^\lambda+\beta}<\frac{1}{\beta}\hspace{20pt}\forall \lambda<0.
\end{align}
%Since the scalar measures induced by spectral measures are always finite\cite{Simon:1980}, constant functions are integrable. 
imply that $\frac{e^{\lambda(1-\epsilon)}}{e^\lambda+\beta}$ is dominated by an integrable function. 
Then, the Lebesgue dominated convergence theorem \cite{Simon:1980} implies that
\begin{align}
\text{lim}_{\epsilon\rightarrow 0}\int \frac{e^{\lambda(1-\epsilon)}}{e^\lambda+\beta} \:\langle x| P(d\lambda)|y\rangle= \int \frac{e^\lambda
}{e^\lambda+\beta}\:\langle x| P(d\lambda)|y\rangle\ .
\end{align}
This is sufficient to guarantee 
\begin{align}
\frac{-i}{2}\text{lim}_{\epsilon\rightarrow 0}\int_{-\infty}^\infty \frac{dt}{\text{sinh}(\pi(t-i\epsilon))}\beta^{-it}\langle y| e^{itK}|x\rangle = \langle y|\frac{\Delta}{\Delta+\beta}|x\rangle.
\end{align}

\section{Interchange of $\beta$ integral with $\epsilon\rightarrow 0$ limit.}\label{sec:Pf5}

Finally, we consider the question of interchange of the order of the $\beta$ integral and the $\epsilon\rightarrow 0$ limit in (\ref{ejn}). For concreteness, consider the $m=2$ in (\ref{ejn}):
\begin{align}
Q_2^{(\epsilon)}&=\int_0^\infty\frac{d\beta}{\beta}\text{lim}_{\epsilon\rightarrow 0}\int dt_0 dt_1 dt_2 \beta^{i(t_0+t_1+t_2)}f(t_0)g_{\epsilon}(t_1)f(t_2)\langle x|\Delta^{-it_0}\bm{\delta}\Delta^{-it_1}\bm{\delta}\Delta^{-it_2}|y\rangle,\\
f(t)&=\frac{1}{2\text{cosh}(\pi t)},\\
g_\epsilon(t)&=\frac{i}{4}\biggl[\frac{1}{\text{sinh}(\pi(t-i\epsilon))}+\frac{1}{\text{sinh}(\pi(t+i\epsilon))}\biggr].
\end{align}
Our aim is to justify bringing the limit $\ep\to 0$ out of the integral. The integrals over the $t_i$ can be done explicitly to get
\bea
&&Q_2^{(\epsilon)}=\int_0^\infty d\beta \biggl(\beta^\epsilon X_\epsilon(\beta)-\beta^{1-\epsilon}Y_\epsilon(\beta)\biggr),\\
&&X_\epsilon(\beta)=\langle x| A\bm{\delta}\biggl(\frac{\Delta^{1-\epsilon}}{\Delta+\beta}\biggr)\bm{\delta}A|y\rangle,\\
&&Y_\epsilon(\beta)=\langle x| A\bm{\delta}\biggl(\frac{\Delta^{\epsilon}}{\Delta+\beta}\biggr)\bm{\delta}A|y\rangle.
\eea

In appendix \ref{sec:Pf4} we show that
\begin{align}
\text{lim}_{\epsilon\rightarrow 0}X_\epsilon(\beta)=\langle x | A\bm{\delta}\biggl(\frac{\Delta}{\Delta+\beta}\biggr)\bm{\delta} A|y\rangle=X(\beta),\\
\text{lim}_{\epsilon\rightarrow 0}Y_\epsilon(\beta)=\langle x | A\bm{\delta}\biggl(\frac{1}{\Delta+\beta}\biggr)\bm{\delta} A|y\rangle=Y(\beta).
\end{align}
%We first split $I_\epsilon$ as follows and then define a new bounded self adjoint positive operator $H$
%\begin{align}
%I_\epsilon&=\int_0^1 d\beta\biggl(\beta^\epsilon X_\epsilon(\beta)-\beta^{1-\epsilon}Y_\epsilon(\beta)\biggr)+\int_1^\infty d\beta\biggl(\beta^\epsilon X_\epsilon(\beta)-\beta^{1-\epsilon}Y_\epsilon(\beta)\biggr),\\
%\frac{\Delta}{\Delta+\beta}&=\int_{\lambda>0}\frac{\lambda}{\lambda+\beta}dE_\lambda,\\
%\end{align}

It is instructive to think of the spectral decomposition of the positive operator $\Delta$:
\bea\label{spectral}
\Delta=\int_{\lam\in\mathcal{R}} e^{-\lam} P(d\lam),
\eea
where $P(d\lambda)$ is a positive-operator valued measure. Then, for any $0<\epsilon<1$, $\beta>0$ and all vectors $|x\rangle$ we have
\bea
&&\langle x|\frac{\Delta^{1-\epsilon}}{\Delta+\beta}|x\rangle\leq\langle x|H|x\rangle
\eea
where
\bea
&&H=\int_{0<\lam<1}\frac{1}{\lam+\beta}P(d\lam)+\int_{\lam>1}\frac{\lam}{\lam+\beta}P(d\lam)\ .
\eea

%If we split the $\beta$ integral 
%It is then easy to check from the definition
%
%Finally, it follows from the Cauchy-Schwarz inequality that for any positive self adjoint bounded operator $Z$, we have
%\begin{align}
%|\langle x |Z|y\rangle|<\langle x|Z|x\rangle+\langle y|Z|y\rangle,\hspace{10pt}\forall x,y\in\mathcal{H}.\label{inequality1}
%\end{align}
By the Cauchy-Schwarz inequality we have the estimates 
\begin{align}
\biggl\vert\beta^\epsilon X_\epsilon(\beta)-\beta^{1-\epsilon}Y_\epsilon(\beta)\biggr\vert&< 2F(\beta),\hspace{20pt}\forall 0<\beta<1,\label{Cond1}\\
\biggl\vert\beta^\epsilon X_\epsilon(\beta)-\beta^{1-\epsilon}Y_\epsilon(\beta)\biggr\vert&< 2\beta F(\beta),\hspace{20pt}\forall 1<\beta,\label{Cond2}\\
F(\beta)&=\langle x | A\bm{\delta} H \bm{\delta} A|x\rangle+\langle y | A\bm{\delta} H \bm{\delta} A|y\rangle.\label{Cond3}
\end{align}

Since $\int_0^1 F(\beta) <\infty$ and $\int_1^\infty \beta F(\beta) <\infty$ the dominated convergence theorem guarantees
\begin{align}
\text{lim}_{\epsilon\rightarrow 0} Q_2^{(\epsilon)} = \int_0^\infty d\beta (X_\beta-\beta Y(\beta)),
\end{align}
which implies
\begin{align}
&\int_0^\infty\frac{d\beta}{\beta}\text{lim}_{\epsilon\rightarrow 0}\int dt_0 dt_1 dt_2\beta^{i(t_0+t_1+t_2)} f(t_0)g_{\epsilon}(t_1)f(t_2)\langle x|\Delta^{-it_0}\bm{\delta}\Delta^{-it_1}\bm{\delta}\Delta^{-it_2}|y\rangle\\&=\text{lim}_{\epsilon\rightarrow 0}\int_0^\infty\frac{d\beta}{\beta}\int dt_0 dt_1 dt_2 \beta^{i(t_0+t_1+t_2)}f(t_0)g_{\epsilon}(t_1)f(t_2)\langle x|\Delta^{-it_0}\bm{\delta}\Delta^{-it_1}\bm{\delta}\Delta^{-it_2}|y\rangle.
\end{align}
The argument above generalizes to the $m^{th}$ term in (\ref{ejn}) and justifies the interchange of the $\beta$ and $\ep\to 0$ limits. This generalization makes it clear that the order of limits of the $\epsilon_i\to 0$ does not matter in (\ref{ejn}).

\bibliography{perturbation}

\begin{thebibliography}{11}%
\makeatletter
\providecommand \@ifxundefined [1]{%
 \@ifx{#1\undefined}
}%
\providecommand \@ifnum [1]{%
 \ifnum #1\expandafter \@firstoftwo
 \else \expandafter \@secondoftwo
 \fi
}%
\providecommand \@ifx [1]{%
 \ifx #1\expandafter \@firstoftwo
 \else \expandafter \@secondoftwo
 \fi
}%
\providecommand \natexlab [1]{#1}%
\providecommand \enquote  [1]{``#1''}%
\providecommand \bibnamefont  [1]{#1}%
\providecommand \bibfnamefont [1]{#1}%
\providecommand \citenamefont [1]{#1}%
\providecommand \href@noop [0]{\@secondoftwo}%
\providecommand \href [0]{\begingroup \@sanitize@url \@href}%
\providecommand \@href[1]{\@@startlink{#1}\@@href}%
\providecommand \@@href[1]{\endgroup#1\@@endlink}%
\providecommand \@sanitize@url [0]{\catcode `\\12\catcode `\$12\catcode
  `\&12\catcode `\#12\catcode `\^12\catcode `\_12\catcode `\%12\relax}%
\providecommand \@@startlink[1]{}%
\providecommand \@@endlink[0]{}%
\providecommand \url  [0]{\begingroup\@sanitize@url \@url }%
\providecommand \@url [1]{\endgroup\@href {#1}{\urlprefix }}%
\providecommand \urlprefix  [0]{URL }%
\providecommand \Eprint [0]{\href }%
\providecommand \doibase [0]{http://dx.doi.org/}%
\providecommand \selectlanguage [0]{\@gobble}%
\providecommand \bibinfo  [0]{\@secondoftwo}%
\providecommand \bibfield  [0]{\@secondoftwo}%
\providecommand \translation [1]{[#1]}%
\providecommand \BibitemOpen [0]{}%
\providecommand \bibitemStop [0]{}%
\providecommand \bibitemNoStop [0]{.\EOS\space}%
\providecommand \EOS [0]{\spacefactor3000\relax}%
\providecommand \BibitemShut  [1]{\csname bibitem#1\endcsname}%
\let\auto@bib@innerbib\@empty
%</preamble>
\bibitem [{\citenamefont {Komatsu}(1966)}]{Komatsu:1966}%
  \BibitemOpen
  \bibfield  {author} {\bibinfo {author} {\bibfnamefont {H.}~\bibnamefont
  {Komatsu}},\ }\href@noop {} {\bibfield  {journal} {\bibinfo  {journal}
  {Pacific Journal of Mathematics}\ }\textbf {\bibinfo {volume} {19}},\
  \bibinfo {pages} {285} (\bibinfo {year} {1966})}\BibitemShut {NoStop}%
\bibitem [{\citenamefont {Conway}\ and\ \citenamefont
  {Morrel}(1987)}]{logarithmop:1987}%
  \BibitemOpen
  \bibfield  {author} {\bibinfo {author} {\bibfnamefont {J.}~\bibnamefont
  {Conway}}\ and\ \bibinfo {author} {\bibfnamefont {B.~B.}\ \bibnamefont
  {Morrel}},\ }\href@noop {} {\bibfield  {journal} {\bibinfo  {journal}
  {Journal of Functional analysis}\ }\textbf {\bibinfo {volume} {70}},\
  \bibinfo {pages} {171} (\bibinfo {year} {1987})}\BibitemShut {NoStop}%
\bibitem [{\citenamefont {Varadarajan}(2013)}]{Liealgebrabook:2013}%
  \BibitemOpen
  \bibfield  {author} {\bibinfo {author} {\bibfnamefont {S.}~\bibnamefont
  {Varadarajan}, \bibfnamefont {Veeravalli}},\ }\href@noop {} {\emph {\bibinfo
  {title} {{Lie Groups, Lie Algebras and their representations}}}}\ (\bibinfo
  {publisher} {Springer Science and Business Media},\ \bibinfo {year}
  {2013})\BibitemShut {NoStop}%
\bibitem [{\citenamefont {Faulkner}(2015)}]{Faulkner:1}%
  \BibitemOpen
  \bibfield  {author} {\bibinfo {author} {\bibfnamefont {T.}~\bibnamefont
  {Faulkner}},\ }\href {\doibase 10.1007/JHEP05(2015)033} {\bibfield  {journal}
  {\bibinfo  {journal} {JHEP}\ }\textbf {\bibinfo {volume} {05}},\ \bibinfo
  {pages} {033} (\bibinfo {year} {2015})},\ \Eprint
  {http://arxiv.org/abs/1412.5648} {arXiv:1412.5648 [hep-th]} \BibitemShut
  {NoStop}%
%%CITATION = ARXIV:1412.5648;%%
\bibitem [{\citenamefont {Balakrishnan}\ \emph {et~al.}(2017)\citenamefont
  {Balakrishnan}, \citenamefont {Faulkner}, \citenamefont {Khandker},\ and\
  \citenamefont {Wang}}]{Faulkner:2}%
  \BibitemOpen
  \bibfield  {author} {\bibinfo {author} {\bibfnamefont {S.}~\bibnamefont
  {Balakrishnan}}, \bibinfo {author} {\bibfnamefont {T.}~\bibnamefont
  {Faulkner}}, \bibinfo {author} {\bibfnamefont {Z.~U.}\ \bibnamefont
  {Khandker}}, \ and\ \bibinfo {author} {\bibfnamefont {H.}~\bibnamefont
  {Wang}},\ }\href@noop {} {\  (\bibinfo {year} {2017})},\ \Eprint
  {http://arxiv.org/abs/1706.09432} {arXiv:1706.09432 [hep-th]} \BibitemShut
  {NoStop}%
%%CITATION = ARXIV:1706.09432;%%
\bibitem [{\citenamefont {Faulkner}\ \emph
  {et~al.}(2016{\natexlab{a}})\citenamefont {Faulkner}, \citenamefont {Leigh},
  \citenamefont {Parrikar},\ and\ \citenamefont {Wang}}]{Faulkner:3}%
  \BibitemOpen
  \bibfield  {author} {\bibinfo {author} {\bibfnamefont {T.}~\bibnamefont
  {Faulkner}}, \bibinfo {author} {\bibfnamefont {R.~G.}\ \bibnamefont {Leigh}},
  \bibinfo {author} {\bibfnamefont {O.}~\bibnamefont {Parrikar}}, \ and\
  \bibinfo {author} {\bibfnamefont {H.}~\bibnamefont {Wang}},\ }\href {\doibase
  10.1007/JHEP09(2016)038} {\bibfield  {journal} {\bibinfo  {journal} {JHEP}\
  }\textbf {\bibinfo {volume} {09}},\ \bibinfo {pages} {038} (\bibinfo {year}
  {2016}{\natexlab{a}})},\ \Eprint {http://arxiv.org/abs/1605.08072}
  {arXiv:1605.08072 [hep-th]} \BibitemShut {NoStop}%
%%CITATION = ARXIV:1605.08072;%%
\bibitem [{\citenamefont {Faulkner}\ \emph
  {et~al.}(2016{\natexlab{b}})\citenamefont {Faulkner}, \citenamefont {Leigh},\
  and\ \citenamefont {Parrikar}}]{Faulkner:4}%
  \BibitemOpen
  \bibfield  {author} {\bibinfo {author} {\bibfnamefont {T.}~\bibnamefont
  {Faulkner}}, \bibinfo {author} {\bibfnamefont {R.~G.}\ \bibnamefont {Leigh}},
  \ and\ \bibinfo {author} {\bibfnamefont {O.}~\bibnamefont {Parrikar}},\
  }\href {\doibase 10.1007/JHEP04(2016)088} {\bibfield  {journal} {\bibinfo
  {journal} {JHEP}\ }\textbf {\bibinfo {volume} {04}},\ \bibinfo {pages} {088}
  (\bibinfo {year} {2016}{\natexlab{b}})},\ \Eprint
  {http://arxiv.org/abs/1511.05179} {arXiv:1511.05179 [hep-th]} \BibitemShut
  {NoStop}%
%%CITATION = ARXIV:1511.05179;%%
\bibitem [{\citenamefont {Faulkner}\ \emph {et~al.}(2017)\citenamefont
  {Faulkner}, \citenamefont {Haehl}, \citenamefont {Hijano}, \citenamefont
  {Parrikar}, \citenamefont {Rabideau},\ and\ \citenamefont
  {Van~Raamsdonk}}]{Faulkner:5}%
  \BibitemOpen
  \bibfield  {author} {\bibinfo {author} {\bibfnamefont {T.}~\bibnamefont
  {Faulkner}}, \bibinfo {author} {\bibfnamefont {F.~M.}\ \bibnamefont {Haehl}},
  \bibinfo {author} {\bibfnamefont {E.}~\bibnamefont {Hijano}}, \bibinfo
  {author} {\bibfnamefont {O.}~\bibnamefont {Parrikar}}, \bibinfo {author}
  {\bibfnamefont {C.}~\bibnamefont {Rabideau}}, \ and\ \bibinfo {author}
  {\bibfnamefont {M.}~\bibnamefont {Van~Raamsdonk}},\ }\href {\doibase
  10.1007/JHEP08(2017)057} {\bibfield  {journal} {\bibinfo  {journal} {JHEP}\
  }\textbf {\bibinfo {volume} {08}},\ \bibinfo {pages} {057} (\bibinfo {year}
  {2017})},\ \Eprint {http://arxiv.org/abs/1705.03026} {arXiv:1705.03026
  [hep-th]} \BibitemShut {NoStop}%
%%CITATION = ARXIV:1705.03026;%%
\bibitem [{\citenamefont {Sárosi}\ and\ \citenamefont
  {Ugajin}(2018)}]{Sarosi:2017}%
  \BibitemOpen
  \bibfield  {author} {\bibinfo {author} {\bibfnamefont {G.}~\bibnamefont
  {Sárosi}}\ and\ \bibinfo {author} {\bibfnamefont {T.}~\bibnamefont
  {Ugajin}},\ }\href {\doibase 10.1007/JHEP01(2018)012} {\bibfield  {journal}
  {\bibinfo  {journal} {JHEP}\ }\textbf {\bibinfo {volume} {01}},\ \bibinfo
  {pages} {012} (\bibinfo {year} {2018})},\ \Eprint
  {http://arxiv.org/abs/1705.01486} {arXiv:1705.01486 [hep-th]} \BibitemShut
  {NoStop}%
%%CITATION = ARXIV:1705.01486;%%
\bibitem [{\citenamefont {Blessenohl}\ and\ \citenamefont
  {Laue}(1988)}]{Blessenohl:1988}%
  \BibitemOpen
  \bibfield  {author} {\bibinfo {author} {\bibfnamefont {D.}~\bibnamefont
  {Blessenohl}}\ and\ \bibinfo {author} {\bibfnamefont {H.}~\bibnamefont
  {Laue}},\ }\href@noop {} {\bibfield  {journal} {\bibinfo  {journal} {Note di
  Matematica}\ }\textbf {\bibinfo {volume} {08}},\ \bibinfo {pages} {111}
  (\bibinfo {year} {1988})}\BibitemShut {NoStop}%
\bibitem [{\citenamefont {Reed}\ and\ \citenamefont
  {Simon}(1980)}]{Simon:1980}%
  \BibitemOpen
  \bibfield  {author} {\bibinfo {author} {\bibfnamefont {M.}~\bibnamefont
  {Reed}}\ and\ \bibinfo {author} {\bibfnamefont {B.}~\bibnamefont {Simon}},\
  }\href@noop {} {\emph {\bibinfo {title} {{Methods of Modern Mathematical
  Physics I: Functional Analysis}}}}\ (\bibinfo  {publisher} {Academic Press},\
  \bibinfo {year} {1980})\BibitemShut {NoStop}%
\end{thebibliography}%
\end{document}